\documentclass[showpacs,amsmath,amssymb,twocolumn,superscriptaddress,notitlepage,preprintnumbers,pra,longbibliography]{revtex4-1}

\usepackage[dvips]{graphicx}
\usepackage{amsmath,amssymb,amsthm,mathrsfs,amsfonts,dsfont}
 
\usepackage{dcolumn}
\usepackage{bm}
\usepackage[english]{babel}
\usepackage{amsmath}
\usepackage{amssymb}
\usepackage{braket}
\usepackage{physics}
\usepackage{amsthm}
\usepackage{natbib}
\usepackage{mathtools}
\usepackage{diagbox}
\usepackage[utf8]{inputenc}
\usepackage[english]{babel}
\usepackage{scalerel}[2014/03/10]
\usepackage{stackengine}
\usepackage{algorithm}
\usepackage{algpseudocode} 
\usepackage{color}
\usepackage{orcidlink}

\usepackage{changes}
\usepackage{multirow}
\usepackage{graphicx} 
\usepackage{subfigure}
\usepackage{relsize}
\usepackage{cprotect}

\usepackage{hyperref}
\hypersetup{colorlinks=true, linkcolor=blue, citecolor=blue, urlcolor=magenta }
 \usepackage{url}

\usepackage{qcircuit}

\graphicspath{{figures/}}

\newcommand{\ud}{\mathrm{d}}
\newcommand{\DD}{\mathcal{D}}

\renewcommand{\qed}{\hfill\blacksquare}
\newcommand{\OO}{\mathcal{O}}

\newcommand{\GG}{\mathcal{G}}

\newcommand{\bos}{\boldsymbol}
\newcommand{\shortxymatrix}[1]{\xymatrix@1@C=.6cm{#1}}
\newcommand{\EZero}{\shortxymatrix{\ar@{-}[r]&}}

\newcommand{\EE}{\mathcal{E}}
\newcommand{\VV}{\mathcal{V}}
\newcommand{\Var}{\mathrm{Var}}

\newcommand{\ii}{\mathbf{i}}

\newtheorem{theorem}{Theorem}

\newtheorem{lemma}{Lemma}

\newtheorem{innercustomthm}{Theorem}
\newenvironment{customthm}[1]
  {\renewcommand\theinnercustomthm{#1}\innercustomthm}
  {\endinnercustomthm}
   
\begin{document}

\title{Imaginary Hamiltonian variational ansatz for combinatorial optimization problems}

\author{Xiaoyang Wang\orcidlink{0000-0002-2667-1879} }

\affiliation{School of Physics, Peking University, Beijing 100871, China
}
\affiliation{Collaborative Innovation Center of Quantum Matter, Beijing 100871, China}
\affiliation{Center for High Energy Physics, Peking University, Beijing 100871, China}
\affiliation{Interdisciplinary Theoretical and Mathematical Sciences Program (iTHEMS), RIKEN, Wako 351-0198, Japan}

\author{Yahui Chai}
\affiliation{Deutsches Elektronen-Synchrotron DESY, Platanenallee 6, 15738 Zeuthen, Germany}

\author{Xu Feng}
\affiliation{School of Physics, Peking University, Beijing 100871, China
}
\affiliation{Collaborative Innovation Center of Quantum Matter, Beijing 100871, China}
\affiliation{Center for High Energy Physics, Peking University, Beijing 100871, China}

\author{Yibin Guo}
\affiliation{Deutsches Elektronen-Synchrotron DESY, Platanenallee 6, 15738 Zeuthen, Germany}

\author{Karl Jansen}
\affiliation{Computation-Based Science and Technology Research Center, The Cyprus Institute, 20 Kavafi Street, 2121 Nicosia, Cyprus}
\affiliation{Deutsches Elektronen-Synchrotron DESY, Platanenallee 6, 15738 Zeuthen, Germany}

\author{Cenk T\"uys\"uz\orcidlink{0000-0003-0257-9784}}
 \affiliation{Deutsches Elektronen-Synchrotron DESY, Platanenallee 6, 15738 Zeuthen, Germany}
\affiliation{Institut für Physik, Humboldt-Universit\"at zu Berlin, Newtonstr. 15, 12489 Berlin, Germany}

\date{\today}
\begin{abstract}
Obtaining exact solutions to combinatorial optimization problems using classical computing is computationally expensive. The current tenet in the field is that quantum computers can address these problems more efficiently. While promising algorithms require fault-tolerant quantum hardware, variational algorithms have emerged as viable candidates for near-term devices. The success of these algorithms hinges on multiple factors, with the design of the ansatz having the utmost importance. It is known that popular approaches such as quantum approximate optimization algorithm (QAOA) and quantum annealing suffer from adiabatic bottlenecks, that lead to either larger circuit depth or evolution time. On the other hand, the evolution time of imaginary time evolution is bounded by the inverse energy gap of the Hamiltonian, which is constant for most non-critical physical systems. In this work, we propose imaginary Hamiltonian variational ansatz ($i$HVA) inspired by quantum imaginary time evolution to solve the MaxCut problem. We introduce a tree arrangement of the parametrized quantum gates, enabling the exact solution of arbitrary tree graphs using the one-round $i$HVA. For randomly generated $D$-regular graphs, we numerically demonstrate that the $i$HVA solves the MaxCut problem with a small constant number of rounds and sublinear depth, outperforming QAOA, which requires rounds increasing with the graph size. Furthermore, our ansatz solves MaxCut exactly for graphs with up to 24 nodes and $D \leq 5$, whereas only approximate solutions can be derived by the classical near-optimal Goemans-Williamson algorithm. We validate our simulated results with hardware demonstrations on a graph with 67 nodes. 
\end{abstract}

\maketitle

\section{Introduction}

Many applications of quantum computers involve the preparation of the ground state of a Hamiltonian system in fields such as chemistry~\cite{McArdle_20}, drug design~\cite{Kirsopp_22,Cao2018}, particle physics~\cite{Klco_2018, Klco_2020}, combinatorial optimization~\cite{Farhi2001Scince,Gemeinhardt2023} and quantum machine learning~\cite{Cerezo_2022QML}. Variational quantum eigensolver (VQE)~\cite{Peruzzo:2014, Kandala2017} is one algorithm designed for ground state preparation on quantum computers. It combines classical optimization techniques with expectation values evaluated on quantum computers. Although VQE has been explored for use on noisy intermediate-scale quantum (NISQ)~\cite{Preskill_2018} devices due to its relatively shallow circuit depth compared to other quantum algorithms, its practical suitability and effectiveness on these devices remain an open question.

The success of VQE highly relies on the efficient parameterization of the quantum circuits. The parametrized quantum circuit is a \textit{variational ansatz} determining what quantum states can be prepared. There have been many efforts to construct the variational ansatz to guarantee that the ground state of a quantum system can be prepared with high accuracy~\cite{Kandala2017,McArdle_20,farhi2014quantum,Wecker_2015}. Among them, the quantum approximate optimization algorithm (QAOA) ansatz is designed to solve combinatorial optimization problems, inspired by the adiabatic evolution~\cite{Farhi2000QuantumCB}. Its performance has been extensively studied both analytically and numerically~\cite{farhi2014quantum,crooks2018performance,Bravyi_2020,farhi2020quantum,Wurtz_2021,BassoJoao2022,Vijendran_2024,zhu2022adaptivequantumapproximateoptimization}. For many-body quantum systems, a widely used ansatz following the same spirit of QAOA is named the Hamiltonian variational ansatz (HVA)~\cite{Wecker_2015}. 


Many challenges exist for the QAOA ansatz. It has been shown that the number of QAOA ansatz rounds should grow linearly to the system size even in some classically solvable tasks to find the solution with high accuracy~\cite{mbeng2019quantum}, and there exists a fundamental limitation if the rounds do not increase faster than a logarithmic function of the system size~\cite{farhi2020quantum}. This requirement leads to other caveats related to the variational optimization of the QAOA ansatz. For example, the ansatz with many rounds is susceptible to noise in NISQ devices~\cite{doi:10.1126/science.abo6587,Pelofske_2024}, and its energy landscape has many local minima~\cite{Zhenduo2023}. More importantly, generic variational ans\"atze with linearly increasing rounds suffer from the barren plateau (BP) phenomenon~\cite{McClean_18,ragone2023unified,fontana2023adjoint}, as demonstrated in Ref.~\cite{Cerezo_21}, so that the gradient of the QAOA ansatz cannot be measured efficiently if its number of rounds grows linearly with the system size. 

The linear behavior originates from the real-time adiabatic evolution that inspires the QAOA ansatz. The adiabatic evolution should be slow enough to avoid diabatic excitation~\cite{bode_adiabatic_2024}, leading to the requirement of many rounds of the QAOA ansatz~\cite{Farhi2000QuantumCB}. There are many efforts to enhance the QAOA ansatz using ideas of, e.g., shortcuts to adiabaticity~\cite{RevModPhys.91.045001}. However, the variational ansatz with high-order counterdiabatic terms contains many unitary gates and is difficult to implement on NISQ devices~\cite{Wurtz2022counterdiabaticity,PhysRevA.105.042415,guan2023singlelayerdigitizedcounterdiabaticquantumoptimization,morris2024performantneartermquantumcombinatorial}.

Imaginary Hamiltonian variational ansatz ($i$HVA), inspired by works of quantum imaginary time evolution (QITE)~\cite{McArdle_19, Motta_20, Sun_21}, is distinguished from the QAOA ansatz. Imaginary time evolution has no problem of diabatic excitation and is widely used in state preparation algorithms such as the tensor networks~\cite{PhysRevLett.91.147902,PhysRevLett.93.040502} and Monte-Carlo~\cite{Gattringer_2014}. $i$HVA has been applied to the Gibbs state preparation in previous studies~\cite{Wang23b}. In this work, we propose to tackle the ground-state problems using $i$HVA. The ansatz uses unitary gates constrained by system symmetries as building blocks, which can be realized on gate-based quantum devices. In this work, we apply $i$HVA to the combinatorial optimization MaxCut problem.

For the MaxCut problem, the arrangement of the parametrized quantum gates in $i$HVA impacts the solution accuracy. We propose a tree arrangement of gates in $i$HVA for arbitrary graphs, and the corresponding ansatz is called the $i$HVA-tree. We provide a theorem which states that arbitrary tree graphs can be maximally cut exactly using $i$HVA-tree with one round and sublinear depth, which cannot be achieved using the constant-round QAOA ansatz~\cite{mbeng2019quantum}. For more complicated random $D$-regular graphs, we perform numerical simulations using noiseless quantum simulators. The results show that $i$HVA-tree can solve the MaxCut of 3-regular graphs exactly up to $14$ graph nodes using constant rounds and sublinear depth, while the QAOA ansatz requires rounds growing with the graph nodes. For $D$-regular graphs with the number of nodes up to $24$ and $D\leq 5$, the two-round $i$HVA-tree can exactly solve the MaxCut problem, whereas only approximate solution can be derived by the classical polynomial-time Goemans-Williamson (G-W) algorithm~\cite{GoemansWilliamson1995,crooks2018performance}. Furthermore, we validate our results on a real quantum device by running an instance of a graph with $67$ nodes.

We show that the constant-round $i$HVA on $D$-regular graphs does not exhibit BPs. It is known that circuits of constant-depth are free from BPs and can be trained efficiently for local Hamiltonians~\cite{Cerezo_21}. For $D$-regular graphs, the constant-round $i$HVA has linear or sublinear depth, where the previous results cannot be applied directly. By exploring the feature that the number of non-commuting gates acting on each qubit has no dependence on the system size, we prove that the variance of the constant-round $i$HVA does not decay exponentially with the graph nodes. Therefore, we prove that the constant-round $i$HVA of $D$-regular graphs is free from BPs.

The remainder of this paper is structured as follows. In Sec.~\ref{sec:Variational ansatz design}, we present how to choose parametrized quantum gates in $i$HVA by leveraging system symmetries and an introduction to the MaxCut problem. In Sec.~\ref{$i$HVA on MaxCut problems}, we explicitly construct $i$HVA for MaxCut following the tree-arrangement. In Sec.~\ref{sec:Numerical results}, numerical simulations are performed to compare the performance of QAOA, $i$HVA and the G-W algorithm. We demonstrate that the constant-round $i$HVA is free from BPs in Sec.~\ref{sec:Trainability of the constant-round $i$HVA}. Finally, in Sec.~\ref{sec:conclusion}, we summarize our results and propose some open questions to be explored in future works.

\section{Framework}\label{sec:Variational ansatz design}
In this section, we review the construction of $i$HVA proposed in Ref.~\cite{Wang23b}. Then, we introduce the combinatorial optimization MaxCut problem and basic concepts of graphs that are used in the following sections.

\subsection{Imaginary Hamiltonian variational ansatz}
Imaginary Hamiltonian variational ansatz is inspired by the QITE algorithm~\cite{McArdle_19, Motta_20, Sun_21}. QITE performs imaginary time evolution on quantum computers with no need for ancillary qubits. Consider a $k$-local Hamiltonian
\begin{align}
    H=\sum_{\mu} H_{\mu},
\end{align}
where $H_{\mu}$ is a local interaction term acting on at most $k$ qubits. The imaginary time propagator of $H$ can be decomposed by a Trotterized-type formula
\begin{equation}
\begin{aligned}
    e^{-\tau H}\ket{\psi}&=(e^{-\Delta\tau H})^L\ket{\psi}\\
    &=(\prod_{\mu} e^{-\Delta\tau H_{\mu}})^L\ket{\psi}+\OO(\frac{\tau^2}{L}),
    \label{eq:Trotter}
\end{aligned}
\end{equation}
where $\Delta\tau=\tau/L$ is the Trotter step. The resulting state approaches the ground state of $H$ when $\tau$ is larger than the inverse energy gap of $H$~\cite{Gattringer_2014}. The inverse of the energy gap typically remains constant with system size for many non-critical physical systems, such as the classical Ising chain~\cite{Farhi2000QuantumCB}. Combinatorial optimization problems, including the MaxCut problem, are often modeled using the classical Ising model. Thus one can expect that the imaginary time evolution converges fast to the ground state in these cases.

For each local interaction term $H_{\mu}$ supported on a set of qubits $\mathbb{S}_{\mu}$, the QITE algorithm~\cite{Motta_20} shows that the imaginary time propagator of each local interaction term can be approximated by unitary gates.
\begin{align}
    e^{-\Delta \tau H_{\mu}}\ket{\psi}\propto \prod_{m\in \mathbf{P}_{\mathbb{S}_i}} e^{-i\theta_{\mu}^{(m)} \sigma_{\mu}^{(m)}/2}\ket{\psi},
    \label{eq:unitary-approximate-qite}
\end{align}
where $\mathbf{P}_{\mathbb{S}_{\mu}}$ includes linear combinations of Pauli strings on the support $\mathbb{S}_{\mu}$ except for identity. For example, if $\mathbb{S}_{\mu}$ includes two qubits, then
\begin{align}
    \sigma_{{\mu}}^{(m)} \in \mathrm{span}(\{IX, IY, IZ,\ldots ZY,ZZ\})
    \label{eq:pauli-expansion}
\end{align}
with real spanning coefficients. We call $\sigma_{\mu}^{(m)}$ a \textit{Pauli series} on $\mathbb{S}_{\mu}$. Eq.~\eqref{eq:unitary-approximate-qite} is approximately valid in the case that the correlation length of the initial state $\ket{\psi}$ and the imaginary time $\Delta \tau$ are finite and not very large~\cite{Motta_20}.

\begin{figure*}[!th]
    \centering
    \includegraphics[width=0.98\linewidth]{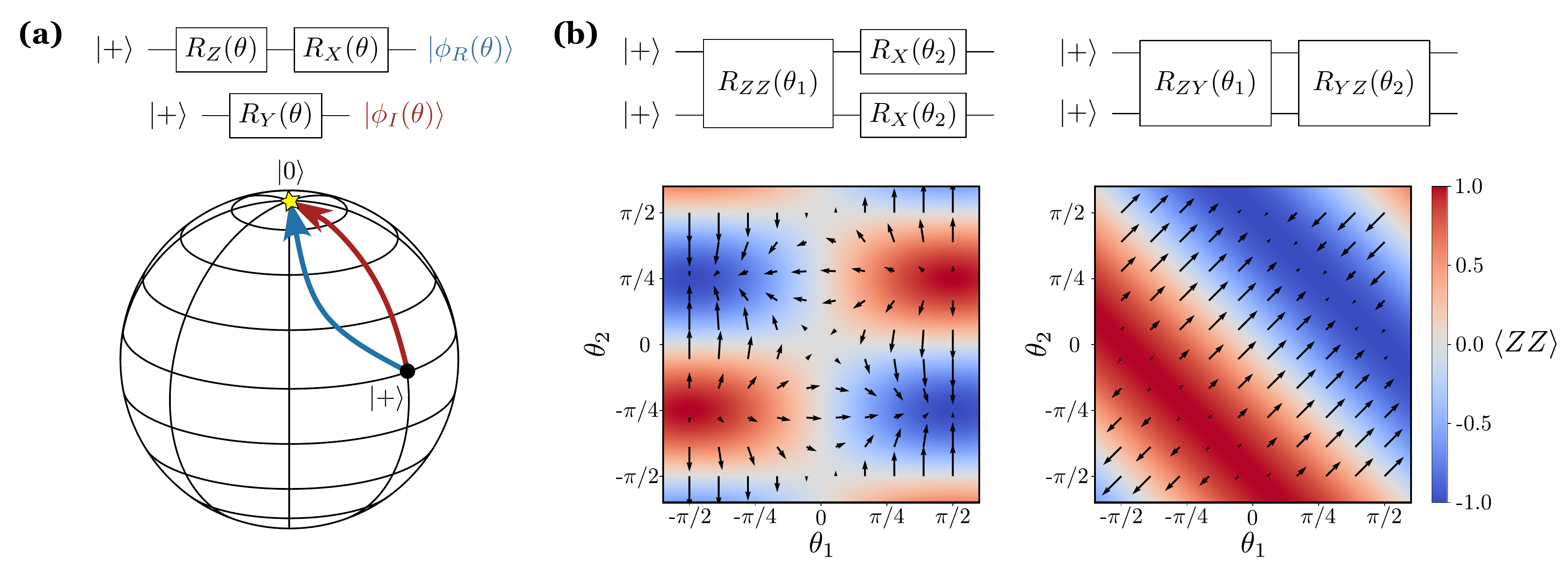}
    \caption{Comparison of $i$HVA and QAOA ansatz finding the ground state of (\textbf{a}) $H_1=-Z$, and (\textbf{b}) $H_2=ZZ$. (\textbf{a}) Gradient descent trajectories of $i$HVA(red) and QAOA(blue) on the Bloch sphere starting at $\theta=0$. (\textbf{b}) The energy landscape of QAOA(left) and $i$HVA(right). The arrows indicate trajectories of the gradient descent.}
    \label{fig:Z-ZZ-examples}
\end{figure*}

Since the imaginary time propagator preserves symmetries of the Hamiltonian system, Pauli series $\sigma_{\mu}^{(m)}$ should also preserve symmetries and thus can be determined. Specifically, assuming $\GG$ is a unitary symmetry group, $H_{\mu}$ and $\ket{\psi}$ are invariant under transformations of the symmetry group, i.e.,
\begin{align}
    [U_g,H_{\mu}]=0, \quad U_g\ket{\psi}=e^{if(g)}\ket{\psi},\forall g\in \GG,
    \label{eq:H_m psi invariance}
\end{align}
where $U_g$ is a unitary representation of the symmetry group element $g\in \GG$, and $f(g)$ is a real scalar function. Then $\sigma_{\mu}^{(m)}$ should be invariant by the conjugation of $U_g$~\cite{Wang23b,Larocca_22, Meyer_22, Sauvage_22, Nguyen_22, Ragone_22}
\begin{align}
   U_g \sigma_{\mu}^{(m)} U_g^{\dagger}=\sigma_{\mu}^{(m)}, \forall g\in \GG.
   \label{eq:symmetry-constrain}
\end{align}
This equation can be solved either by constructing linear systems of equations~\cite{Wang23b} or by implementing twirling operations on $\sigma_{\mu}^{(m)}$~\cite{Ragone_22}. Then, $i$HVA is constructed by applying unitary gates in Eq.~\eqref{eq:unitary-approximate-qite} for each local interaction term 
\begin{align}
    \ket{\phi_I^{(L)}(\bos{\theta})}=\prod_{l=1}^{L}\prod_m \prod_{\mu} e^{-i\theta_{l,\mu}^{(m)}\sigma_{\mu}^{(m)}/2},
    \label{eq:QITE-inspired-ansatz}
\end{align}
where $L$ is the number of ansatz layers. $\mu,m$ index local interaction terms and symmetry preserving Pauli series, respectively. 

In this work, we compare the $i$HVA and the QAOA ansatz~\cite{farhi2014quantum, Wecker_2015}. The QAOA ansatz encodes real-time evolution $e^{-itH}$ of the Hamiltonian $H$. Since the real-time propagator also preserves unitary symmetries of the Hamiltonian system, the Pauli series utilized in QAOA ansatz also satisfy Eq.~\eqref{eq:symmetry-constrain}. Thus, the Pauli series obtained by Eq.~\eqref{eq:symmetry-constrain} include the ones in QAOA if only unitary symmetry groups are considered. These two kinds of ans\"atze are distinguished if the Hamiltonian system possesses anti-unitary time-reversal symmetry, which means that the Hamiltonian and the initial state have only purely real entries. Time-reversal symmetry is preserved by many chemical, quantum field, and combinatorial optimization Hamiltonians, such as the MaxCut problem studied in this work. For these Hamiltonian systems, the Pauli series used in $i$HVA and QAOA ansatz are distinguished by
\begin{align}
    \sigma_{\mu}^{(m)}: \left\{ \begin{array}{ll}
    \text{contains odd $Y$ letters,} & \text{for $i$HVA;}\\
 \text{contains even $Y$ letters,} & \text{for QAOA.}
  \end{array} \right.
  \label{eq:even-odd-requirement}
\end{align}
For example, in the two-qubit case, $\sigma_{\mu}^{(m)}$ of the $i$HVA is spanned by
\begin{align}
    \sigma_{\mu}^{(m)}\in \mathrm{span}(\{IY,XY,YI,YX,YZ,ZY\}),
\end{align}
and of the QAOA ansatz
\begin{align}
    \sigma_{\mu}^{(m)}\in \mathrm{span}(\{IX,IZ, XI, XX, XZ,YY, ZI, ZX,ZZ\}).\nonumber 
\end{align}

The discriminative criterion Eq.~\eqref{eq:even-odd-requirement} follows intuitions. For $i$HVA, $\sigma_{\mu}^{(m)}$ with odd $Y$ letters is purely imaginary. Since $H_{\mu}$ is purely real, $e^{-i\theta\sigma_{\mu}^{(m)}/2}$ can be regarded as performing real-time dynamics of imaginary Hamiltonian $-iH_{\mu}$, which corresponds to the imaginary time propagator $e^{-\Delta \tau H_i}$. This is the reason we refer to this ansatz as the imaginary Hamiltonian variational ansatz ($i$HVA).  For QAOA, on the other hand, $\sigma_{\mu}^{(m)}$ with even $Y$ letters is purely real, which is consistent with the realness of $H_i$. We denote the Pauli series used in $i$HVA as the \textit{relevant series}. 

To highlight the difference between $i$HVA and QAOA ansatz, we present two toy examples. The variational ansatz state of $i$HVA and QAOA are denoted by $\ket{\phi_I}$ and $\ket{\phi_R}$, respectively. Consider a one-qubit and a two-qubit Hamiltonian
\begin{align}
    H_1=-Z;\quad  H_2=ZZ,
\end{align}
whose $i$HVA and QAOA ans\"atze are
\begin{align}
H_1:&\left\{ \begin{array}{l}
 \ket{\phi_I(\theta)} = R_Y(\theta)\ket{+}\\
 \ket{\phi_R(\theta)} = R_X(\theta)R_Z(\theta)\ket{+}
  \end{array} \right. ;\\
 H_2:  & \left\{ \begin{array}{l}
 \ket{\phi_I(\theta_1,\theta_2)} = R_{YZ}(\theta_2)R_{ZY}(\theta_1)\ket{++}\\
 \ket{\phi_R(\theta_1,\theta_2)} = R_{XI}(\theta_2)R_{IX}(\theta_2)R_{ZZ}(\theta_1)\ket{++}
  \end{array} \right. .\nonumber
\end{align}
where $R_{\sigma}(\theta)=e^{-i\sigma\theta/2}$ is the Pauli exponential of Pauli string $\sigma$. The Pauli strings in $i$HVA and QAOA ansatz contain odd and even $Y$ letters, respectively, and all satisfy the symmetry constraint in Eq.~\eqref{eq:symmetry-constrain} ($U_g=XX$ for the two-qubit case). $i$HVA for these two toy examples are closely related to the imaginary-time evolution of $H_1$ and $H_2$, as one can check that 
\begin{equation}
\begin{aligned}
    e^{\tau Z}\ket{+}&\propto e^{-i\theta(\tau)Y/2}\ket{+};\\
    e^{-\tau ZZ}\ket{++}&\propto e^{-i\theta_1(\tau)ZY/2}\ket{++}=e^{-i\theta_2(\tau)YZ/2}\ket{++},\nonumber
\end{aligned}
\end{equation}
where $\theta(\tau),\theta_1(\tau),\theta_2(\tau)$ are functions of the imaginary time $\tau$. In these formulae, all the imaginary-time propagator and the unitary gates can be represented by real matrices in the computational basis, as a result of the time-reversal symmetry kept by $H_1$ and $H_2$.
\begin{figure*}
    \centering
    \includegraphics[width=0.80\textwidth]{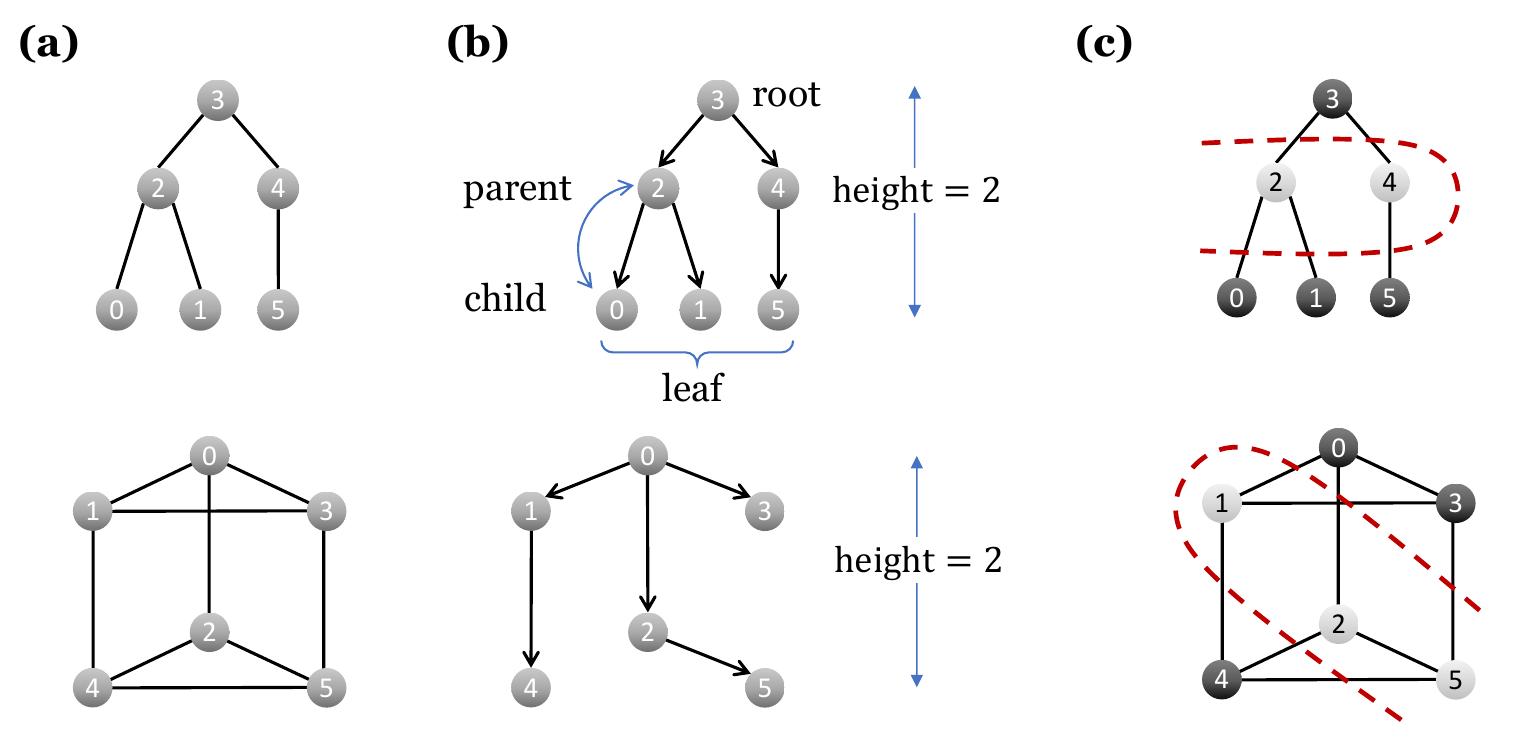}
    \caption{(\textbf{a}) Examples of a tree graph (upper row) and a 3-regular graph (lower row). (\textbf{b}) Oriented spanning trees of the tree graph and the 3-regular graph. Each oriented edge connects a child node at its tip and a parent node at its tail. A root node is the topmost node in an oriented tree that has no parent node, and the leaf node does not have child nodes. The height of an oriented tree is the length of the longest downward path from the unique root node to one of the leaf nodes. (\textbf{c}) The MaxCut solution of the tree graph and the 3-regular graph.}
    \label{fig:graph}
\end{figure*}

We study the ground-state preparation trajectories in these two ans\"atze of the one and two-qubit examples, as presented in Fig.~\ref{fig:Z-ZZ-examples}. Fig.~\ref{fig:Z-ZZ-examples}(\textbf{a}) shows two trajectories in the Bloch sphere as we perform gradient descent using $i$HVA and QAOA of $H_1$ starting at $\theta=0$. We see that the $i$HVA trajectory (red) is geodesic between the initial state ($\ket{+}$) and the ground state of $H_1$ ($\ket{0}$) on the Bloch sphere, while the trajectory of QAOA (blue) is non-geodesic and would require more iteration steps during the gradient descent. Fig.~\ref{fig:Z-ZZ-examples}(\textbf{b}) shows the energy landscapes of $i$HVA and QAOA for $H_2$. We see that a saddle point in the QAOA landscape appears at $(\theta_1,\theta_2)=(0,0)$, which complicates the optimization process. In contrast, the $i$HVA landscape does not have this problem and is thus more favorable for optimization.

\subsection{Graph and oriented spanning tree}

We review the concepts of tree graph and $D$-regular graph, which are the types of graphs mainly studied in this work. A graph $G=(\VV,\EE)$ consists of a set of $N$ nodes $i\in\VV$, labeled by integers $i=0,\ldots, N-1$, and undirected edges $(i,j)\in\EE$. A \textit{tree graph} is defined as a graph without cycle. A graph is called as a $D$\textit{-regular graph} if each node in the graph has $D$ edges connected with the other nodes. Figure~\ref{fig:graph}(\textbf{a}) presents an example of the tree graph with six nodes, five edges, and the 3-regular graph with six nodes, nine edges.

A key concept used in this work is the \textit{oriented spanning tree}, as shown in Fig.~\ref{fig:graph}(\textbf{b}). A spanning tree of an undirected graph $G$ is a tree subgraph that includes all of the vertices of $G$. An oriented spanning tree is obtained by choosing a tree node as the root node, such that the tree hierarchy is subsequently constructed. Each edge of the oriented tree connects a parent node and a child node. For a tree graph, its oriented spanning tree is not unique, which is determined by the chosen root. Oriented spanning trees of a tree graph and a 3-regular graph are shown in Fig.~\ref{fig:graph}(\textbf{b}), and other useful concepts of the oriented tree are provided in the figure. 

\subsection{MaxCut problem}

The MaxCut problem is a paradigmatic test for various ans\"atze used in VQE~\cite{farhi2014quantum}. Given that a graph consists of edges and nodes, MaxCut aims to partition the graph's nodes into two complementary sets, such that the number of edges between these two sets is as large as possible. This problem can be formulated as follows: Suppose $G=(\VV,\EE)$ is a graph with $N$ nodes. Given an $N$-bit string $x=x_{N-1}\ldots x_0,x_i\in\{0,1\}$, assume $\mathrm{cut}(x)$ be the set of edges $(i,j)$ such that $x_{i}\neq x_{j}$. The object of MaxCut is to maximize the cut size $C(x)=|\mathrm{cut}(x)|$, i.e., the number of edges in $\mathrm{cut}(x)$. For example, Fig.~\ref{fig:graph}(\textbf{c}) gives solutions of the MaxCut of the two graphs in Fig.~\ref{fig:graph}(\textbf{a}), where the black (white) node denotes $x_i=1~(0)$. The MaxCut solution of the tree graph is $x_{\mathrm{tree}}=101011$ with cut size $C(x_{\mathrm{tree}})=5$, and of the 3-regular graph is $x_{\mathrm{regular}}=011001$ with cut size $C(x_{\mathrm{regular}})=7$. The MaxCut solution of a graph is not unique. Specifically, the bit-wise inverse $\bar{x}_{\mathrm{tree}}=010100$ and $\bar{x}_{\mathrm{regular}}=100110$ are also MaxCut solutions of their corresponding graphs.

For arbitrary graphs, the MaxCut solution is equal to the maximum eigenvalue of an $N$-qubit Hamiltonian
\begin{align}
    \hat{C}=\frac{1}{2}\sum_{(i,j)\in\EE}(I-Z_i Z_j),
    \label{eq:hat-C}
\end{align}
where $Z_i$ is the Pauli-$Z$ operator on the $i$-th qubit and $I$ is the identity operator. Finding the maximum of $\hat{C}$ is equivalent to finding the minimum of $-\hat{C}$. Thus, we aim to find the minimum eigenvalue of the MaxCut Hamiltonian
\begin{align}
    H_{\mathrm{MC}}=\sum_{(i,j)\in\EE}Z_i Z_j,
\end{align}
where an irrelevant constant is discarded. Thus, the MaxCut problem is mapped to a ground-state problem and can be solved using VQE~\cite{farhi2014quantum,Wurtz_2021}. 

Finding the MaxCut solution for arbitrary graphs is known to be NP-complete~\cite{Karp1972}. For this reason, we intend to find an approximate solution for MaxCut. The performance of an algorithm approximately solving MaxCut can be estimated by the approximation ratio, which is defined by
\begin{align}
    \alpha \equiv\frac{C(x)}{C_{\max}},
    \label{eq:approx-ratio-definition}
\end{align}
where $C_{\max}$ is the exact maximum cut size of the graph. $C(x)$ is the cut size provided by a given algorithm. $\alpha\rightarrow 1$ indicates that the algorithm could solve MaxCut with high accuracy. The Goemans-Williamson algorithm is a classical polynomial-time algorithm that guarantees an approximation ratio of 0.8785~\cite{GoemansWilliamson1995}, which is optimal under the unique games conjecture~\cite{Khot2007}. 

In the quantum scenario, $C(x)$ is transformed to a functional of a given quantum state $\ket{\phi}$, which is a superposition of bit strings $\ket{x}$, and the cut size is evaluated by 
\begin{equation}
\begin{aligned}
    C(x)\rightarrow C[\phi] &\equiv \bra{\phi}\hat{C}\ket{\phi} \\
    &=\frac{1}{2}(|\EE|-\bra{\phi}H_{\mathrm{MC}}\ket{\phi}),
\end{aligned}
\end{equation}
where $\hat{C}$ is defined in Eq.~\eqref{eq:hat-C}, $N$ is the number of nodes, $|\EE|$ is the total number of edges of the graph. $\ket{\phi}$ is the variational ansatz state, which can be either the $i$HVA state $\ket{\phi_I}$ or the QAOA ansatz state $\ket{\phi_R}$ in this work. 

\section{$i$HVA for MaxCut}\label{$i$HVA on MaxCut problems}
This section explicitly constructs the $i$HVA for the MaxCut problem. Apart from providing relevant series of the MaxCut Hamiltonian, we focus on choosing an appropriate arrangement of the parametrized quantum gates, where the tree arrangement and the $i$HVA-tree are introduced for arbitrary graphs.

\subsection{Relevant series of MaxCut}

The relevant series of the MaxCut and the general structure of its $i$HVA are given as follows. $H_{\mathrm{MC}}$ commutes with the symmetry transformation $U_g=\prod_{i\in\VV}X_i$, which corresponds to the global bit-flip symmetry~\cite{Bravyi_2020}. Additionally, $H_{\mathrm{MC}}$ is purely real such that the time-reversal symmetry is preserved. The relevant series corresponding to the local interaction term $Z_iZ_j$ are 
\begin{align}
   \sigma_{(i,j)}^{(1)}= Z_iY_j, \quad \sigma_{(i,j)}^{(2)}= Y_i Z_j.
\end{align}
Then, we define parametrized subcircuits of these two relevant series
\begin{align}
    U^{(l)}_{ZY}\equiv \prod_{(i,j)\in\EE}e^{-i\theta_{l,ij}Z_i Y_j/2},\;U^{(l)}_{YZ}\equiv \prod_{(i,j)\in\EE}e^{-i\theta_{l,ij}Y_i Z_j/2},
    \label{eq:UZY-UYZ}
\end{align}
where $\theta_{l,ij}$ are variational parameters. The variational ansatz is constructed by alternating these subcircuits in order and applying to the initial state 
\begin{align}
\ket{\phi_I^{(p)}(\bos{\theta})}\equiv U^{(p)}_{ZY/YZ} \ldots U^{(2)}_{YZ}U^{(1)}_{ZY}\ket{+}^{\otimes N},
\label{eq:max-cut-qite-ansatz}
\end{align}
where $\ket{+}^{\otimes N}$ is the tensor product of $N$ single-qubit states $\ket{+}=(\ket{0}+\ket{1})/\sqrt{2}$. $U^{(p)}_{ZY/YZ}$ denotes that the last round is $U^{(p)}_{ZY(YZ)}$ if $p$ is odd (even).  Similar to $L$ used in Eq.~\eqref{eq:QITE-inspired-ansatz}, here we define $p$ as the number of rounds of $i$HVA. In the first round of $i$HVA, we apply one $ZY$ gate on each edge of the graph. In the second round, we reverse the qubits of $Z$ and $Y$ such that both $e^{-i\theta_{l,ij}Z_i Y_j/2}$ and $e^{-i\theta_{l,ij}Y_i Z_j/2}$ are applied on one edge, as required by the imaginary-time evolution of $Z_iZ_j$. In one round of $i$HVA, only one $ZY$ gate is applied for each edge of the graph. So in this way, the number of two-qubit gates in one round of $i$HVA can be compared with that in one round of QAOA ansatz, as will be detailed later. 

The QAOA ansatz for the MaxCut problem is distinguished from $i$HVA, which reads
\begin{equation*}
\begin{aligned}
    &\ket{\phi_R^{(p)}(\bos{\beta},\bos{\gamma})}\\
    =&\prod_{l=1}^p \left[\prod_{i\in\VV}e^{-i\beta_{l,i}X_i/2}\prod_{(i,j)\in\EE}e^{-i\gamma_{l,ij}Z_iZ_j/2} \right]\ket{+}^{\otimes N},
\end{aligned}
\end{equation*}
where $\beta_{l,i},\gamma_{l,ij}$ are variational parameters. This ansatz is the multi-angles QAOA (ma-QAOA) ansatz~\cite{Herrman_2022}, which has better expressibility than the original QAOA ansatz~\cite{blekos2023review}. The number of two-qubit gates in one round of the QAOA ansatz is the same as in one  round of $i$HVA, since the two-qubit Pauli exponentials in these two ans\"atze can be converted by single-qubit gates
\begin{align}
    e^{-i\theta_{l,ij}Z_i Y_j/2} = e^{i\frac{\pi}{4}X_j} e^{-i\theta_{l,ij}Z_i Z_j/2} e^{-i\frac{\pi}{4}X_j}.
\end{align}


One round of $i$HVA has $ZY$ gates $e^{-i\theta_{l,ij}Z_i Y_j/2}$ on different edges that do not commute with each other. This allows different \textit{arrangements}. The arrangement of the $ZY$ gates impacts the solution accuracy. Additionally, depending on the arrangement of the $ZY$ gates, the circuit depth of $U^{(l)}_{ZY}$ could be a constant or grow logarithmically or even linearly to the graph nodes $N$. We discuss the arrangement of the $i$HVA and its depth in detail in the following subsections.

\subsection{$i$HVA on trees and tree arrangement}\label{tree arrangement}
In this subsection, we demonstrate how to choose an appropriate arrangement of Pauli exponentials $e^{-i\theta_{l,ij}Z_i Y_j/2}$ in one subcircuit $U^{(l)}_{ZY}$ defined in Eq.~\eqref{eq:UZY-UYZ}. Our choice of the tree arrangement is based on an observation that the MaxCut of tree graphs can be exactly achieved by one round of $i$HVA by choosing the tree arrangement, as demonstrated below.

\begin{figure}[!h]
    \centering
    \includegraphics[width=0.48\textwidth]{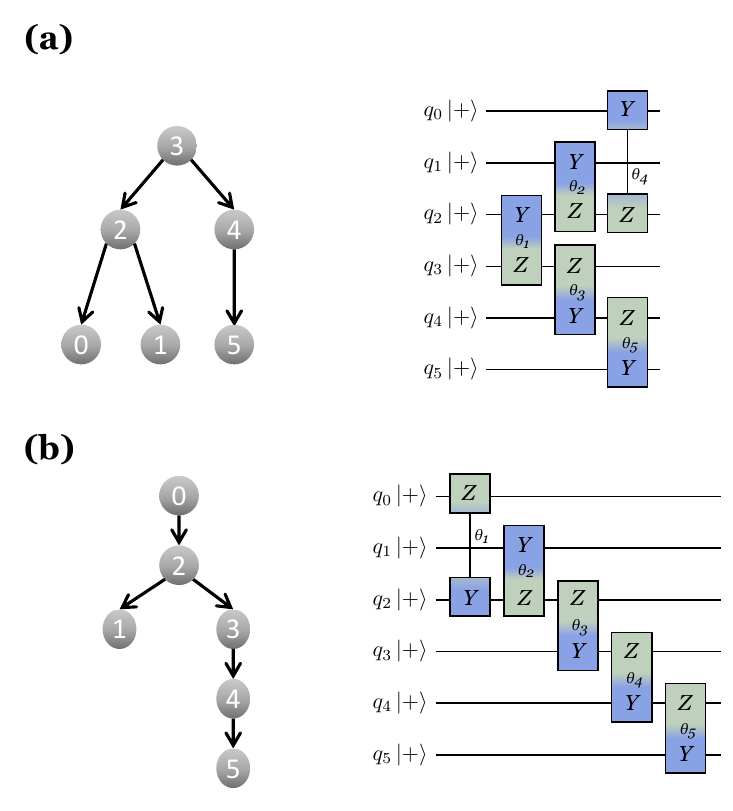}
    \caption{(\textbf{a}) The oriented tree for the tree graph in Fig.~\ref{fig:graph}(\textbf{a}) and the corresponding tree arrangement of the $ZY$ gates. Each colored rectangle is a $ZY$ exponential $e^{-i\theta_{l,ij}Z_iY_j/2}$ with the value of $\theta_{l,ij}$ shown at the center of the rectangle. This oriented tree has the node $3$ as the root, which is the lowest among all orientations of the tree. (\textbf{b}) The highest-oriented tree of the tree graph has node $0$ as the root node. Its corresponding tree arrangement of $ZY$ gates is illustrated.}
    \label{fig:tree-arranegment-on-tree}
\end{figure}

We use the tree graph in Fig.~\ref{fig:graph}(\textbf{a}) as an example, whose MaxCut solution is shown in the upper panel of Fig.~\ref{fig:graph}(\textbf{c}). The two solutions $x_{\mathrm{tree}}=101011$ and $\bar{x}_{\mathrm{tree}}=010100$ can be obtained by preparing the ground state of its MaxCut Hamiltonian $H_{\mathrm{MC}}$
\begin{align}
    \ket{G.S.}=\frac{1}{\sqrt{2}}(\ket{010100}+\ket{101011}).
    \label{eq:tree-ground-state}
\end{align}
This ground state is locally equivalent to the $6$-qubit Greenberger-Horne-Zeilinger(GHZ) state $(\ket{0}^{\otimes 6}+\ket{1}^{\otimes 6})/\sqrt{2}$. GHZ state has long-range entanglement and cannot be prepared by a constant-depth quantum circuit~\cite{Bravyi_2006}. This provides the intuition to arrange the $ZY$ gates following an orientation of the tree to increase the circuit depth. Fig.~\ref{fig:tree-arranegment-on-tree}(\textbf{a}) shows the oriented tree and the corresponding \textit{tree arrangement} of $ZY$ gates in $U^{(l)}_{ZY}$. This circuit with variational parameters $\theta_i, i=1,\ldots, 5$ is a one-round $i$HVA. One can check that the ground state Eq.~\eqref{eq:tree-ground-state} can be prepared by setting
\begin{align}
    \theta_1=\theta_2=\ldots =\theta_{5}&=\frac{\pi}{2}.
\end{align}
Thus, the MaxCut of the tree is exactly solved.

\begin{figure*}
    \centering
    \includegraphics[width=0.95\textwidth]{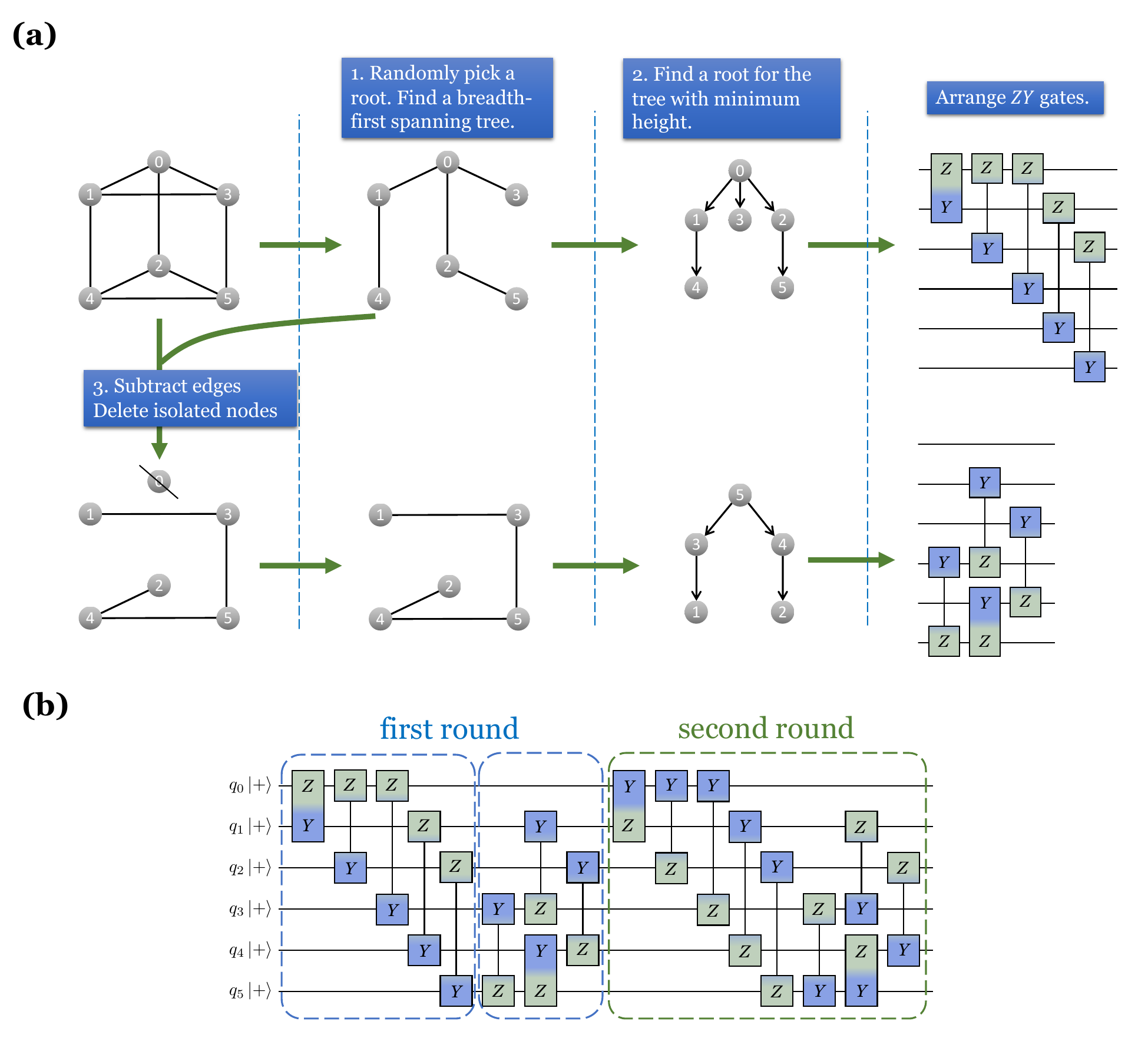}
    \caption{(\textbf{a}) Construction of the $i$HVA-tree solving MaxCut of arbitrary graphs. These three steps give one round $U^{(l)}_{ZY}$ of the $i$HVA-tree. Each two-qubit gate represents one $ZY$ exponential $e^{-i\theta_{l,ij}Z_iY_j/2}$. (\textbf{b}) $i$HVA-tree with two rounds. The first round has two parts that arrange $ZY$ gates by the procedure in (\textbf{a}). The second round is constructed by reversing the orientation of $ZY$ gates to $YZ$.}
    \label{fig:qite-tree-ansatz}
\end{figure*}

This example can be generalized to arbitrary trees, and we have the following theorem.
\begin{theorem}\label{theorem1}
The MaxCut of arbitrary trees can be achieved by the one-round $i$HVA following the tree arrangement.
\end{theorem}
\noindent The proof of this theorem is presented in Appendix~\ref{app:exact_cut_of_tree}.

In the tree arrangement, we construct an oriented tree whose root can be chosen arbitrarily. We can use this arbitrariness to reduce the depth of the tree arrangement ansatz. The reduction in depth reduces the runtime of the algorithm and improves robustness against noise on real hardware. Roughly speaking, the tree arrangement circuit is shallower if the corresponding oriented tree is lower in height. For example, the tree arrangement in Fig.~\ref{fig:tree-arranegment-on-tree}(\textbf{a}) has depth $3$ in the unit of the depth of the $ZY$ exponential, and the height of the oriented tree is $2$. On the other hand, if the node $0$ is chosen as the root, as shown in Fig.~\ref{fig:tree-arranegment-on-tree}(\textbf{b}), the corresponding tree arrangement has depth $5$, and the height of the oriented tree is $4$. Thus, to construct a tree arrangement circuit with relatively small depth, we choose the root node corresponding to \textit{the lowest oriented tree} among all nodes. This can be done on a classical computer by first enumerating all the $N$ nodes as the root and then calculating the corresponding height of the oriented tree. As calculating the height of an oriented tree recursively requires the time complexity of $\OO(N)$~\cite{CalculateHeight}, the above procedure can be accomplished with the time complexity of $\OO(N^2)$.

\subsection{$i$HVA on arbitrary graphs}

In the previous subsection, we have seen that the one-round $i$HVA following the tree arrangement can achieve the MaxCut of arbitrary trees. In this subsection, we generalize the tree arrangement of tree graphs to arbitrary connected graphs. 

The generalization proceeds by decomposing a connected graph into several \textit{breadth-first spanning (BFS) trees}~\cite{AlgorithmDesign2005}. We choose the BFS tree because the BFS tree is usually lower in height than the other spanning trees, so the corresponding quantum circuit has a smaller depth. One way of decomposing a connected graph into BFS trees is shown in Fig.~\ref{fig:qite-tree-ansatz}(\textbf{a}). In the first step, we randomly pick a root and construct a BFS tree, which means that a parent node connects all the adjacent nodes as child nodes if the tree has never visited the nodes. Constructing a BFS tree with a given root is efficient for all connected graphs. In the second step, we find a root of the spanning tree leading to the minimum height, and arrange the $ZY$ gates at the right-most end of the circuit, as described in the previous subsection. Thirdly, apart from the spanning tree, the remaining graph is obtained by subtracting edges in the spanning tree from the original graph, and deleting possibly appeared isolated nodes. We delete the isolated nodes to provide an explicit judgment on when the procedure should be stopped, as will be detailed later. The remaining graph can be connected or disconnected. For every connected part of the remaining graph, we return to the first step and repeatedly construct its BFS tree, as shown in the second line of Fig.~\ref{fig:qite-tree-ansatz}(\textbf{a}). This repetition is stopped in the third step if no nodes are left after deleting isolated nodes. These steps are summarized in the following algorithm. 

\begin{algorithm}[H]
\caption{Arrange gates in one round of $i$HVA-tree}\label{alg:qite-ansatz}
\begin{algorithmic}
 \Require A connected graph $G=(\VV,\EE)$
 \Procedure{arrange}{$G$}
 \State 1. Randomly pick a root of $G$. Construct a breadth-first spanning tree $T$.
 \State 2. Find a root $r$ for the tree with minimum height. Arrange $ZY$ gates following oriented spanning tree $T$ with root $r$.
 \State 3. Define the remaining graph $G' \gets  G-T$.
 \State Delete isolated nodes in $G'$.
  \If{$G'$ has no nodes}
    \State \Return
  \Else
    \For{each connected subgraph $g'$ of $G'$}
    \State \Call{arrange}{$g'$}.
    \EndFor
  \EndIf
\EndProcedure
\end{algorithmic}
\end{algorithm}

This algorithm can be performed efficiently on classical computers. The root-finding procedure in the second step is the most time-consuming part of the algorithm. Assume that the number of edges in each spanning tree derived by the above procedure is $M_{\EE_{\alpha}}$, where $\EE = \bigcup_{\alpha} \EE_{\alpha}$ is the whole edge set of $G$, and $M=\sum_{\alpha} M_{\EE_{\alpha}}$ is the total number of edges of $G$. The number of nodes in each tree is $M_{\EE_{\alpha}}+1$. Thus, using the time complexity~$\OO(N^2)$ of one tree derived in the previous subsection, the total time complexity of the algorithm is upper bounded by 
\begin{align}
    \sum_{\alpha} (M_{\EE_{\alpha}}+1)^2\sim \OO(M^2),
\end{align}
which grows polynomially to the system size.

Using the above procedure, an explicit product order of $ZY$ gates in the $U^{(l)}_{ZY}$ is obtained. The same order can be defined for another subcircuit $U^{(l)}_{YZ}$ by reversing the orientation of $ZY$ gates in $U^{(l)}_{ZY}$ to $YZ$, as shown in Fig.~\ref{fig:qite-tree-ansatz}(\textbf{b}). We call $i$HVA with each round given by the above procedure as \textit{$i$HVA-tree}. In Fig.~\ref{fig:qite-tree-ansatz}(\textbf{b}), we show an example of $i$HVA-tree with two rounds. 

For tree graphs, one-round $i$HVA-tree is reduced to the tree arrangement introduced in Sec.~\ref{tree arrangement}. Thus, one-round $i$HVA-tree can exactly cut arbitrary tree graphs, as shown in Theorem~\ref{theorem1}. In contrast, QAOA requires linearly growing rounds to exactly cut lines~\cite{mbeng2019quantum}, and ma-QAOA with one round can exactly cut only star graphs~\cite{Herrman_2022}. Both lines and star graphs are particular tree graphs and can be cut exactly using the one-round $i$HVA-tree.

The advantage of $i$HVA-tree over QAOA on tree graphs is in exchange for the larger depth of the quantum circuit. Assuming an all-to-all qubit connectivity of the quantum chip, the depth of $p$-round QAOA for $D$-regular graph is $\OO(p)$, which has no dependence on the number of nodes $N$~\cite{Bravyi_2020}. In contrast, the depth of the $p$-round $i$HVA-tree is lower bounded by
\begin{align}
    d_p =\Omega(p\log N),
\end{align}
and upper bounded by
\begin{align}
    d_p =\OO(p N),
\end{align}
which is distinguished from QAOA. These bounds are derived in Appendix~\ref{app:circuit-depth-of-iHVA}. This depth scaling could bring up fundamental differences between the accuracy of solving MaxCut using $i$HVA-tree and QAOA. Although deeper quantum circuits suffer more from errors on NISQ devices, there exist error suppression methods such as dynamical decoupling~\cite{PhysRevA.58.2733,DUAN1999139,PhysRevLett.82.2417}, that are suitable for the $i$HVA-tree type circuits.

The arrangement of the $ZY$ gates in the $i$HVA impacts its ability to find the MaxCut solution. To manifest this, we introduce another arrangement of the $i$HVA in Appendix~\ref{app:$i$HVA-stagger ansatz}, constructed as shallow as possible among all the arrangements by a staggered layout of the $ZY$ gates. $i$HVA following this arrangement is called the \textit{$i$HVA-stagger} ansatz. In the following section, we numerically compare $i$HVA-tree, $i$HVA-stagger, and QAOA ansatz by testing their performance of finding the MaxCut solution of random regular graphs.

\begin{figure*}
\centering
\includegraphics[width=1\textwidth]{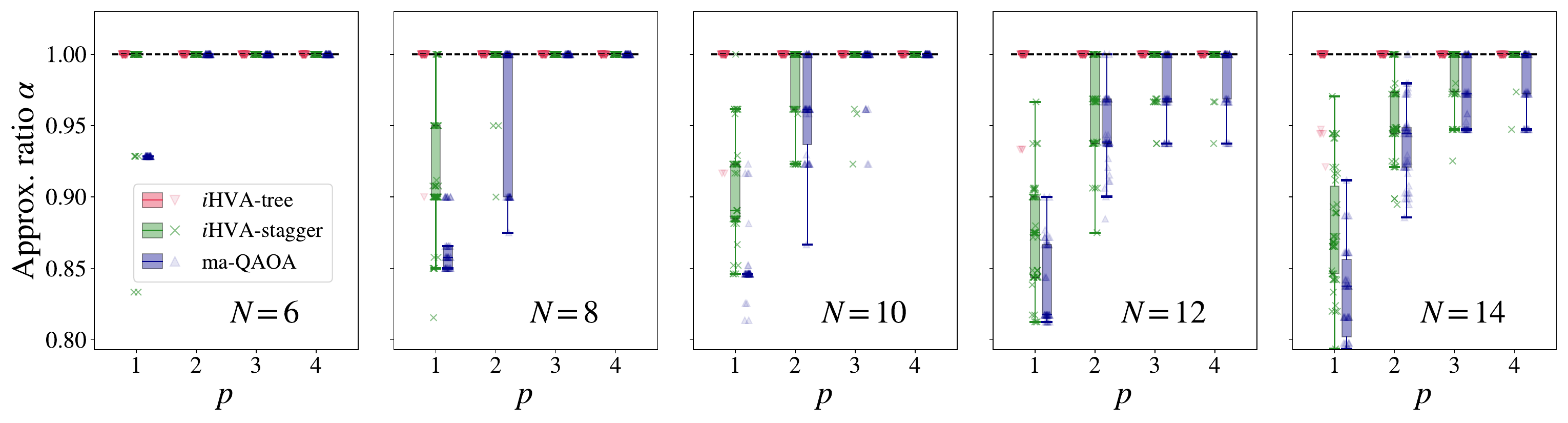}
\caption{Simulated results for the approximation ratio $\alpha$ of 3-regular graphs as a function of $p$ circuit rounds. The ans\"atze considered here include $i$HVA-tree, $i$HVA-stagger, and ma-QAOA, with results marked by red lower-triangle, green cross, and blue upper-triangle, respectively. Each subplot corresponds to a fixed number of nodes $N$, with $50$ randomly generated 3-regular graphs. The box plot is used to reflect the statistical properties of the $50$ ratios. For $N\leq 14$, approximation ratios achieved by $i$HVA-tree are all close to $1$ as $p\geq 2$. }
\label{fig:qite-tree-3-regular}
\end{figure*}   

\begin{figure*}
    \centering
    \includegraphics[width=1\textwidth]{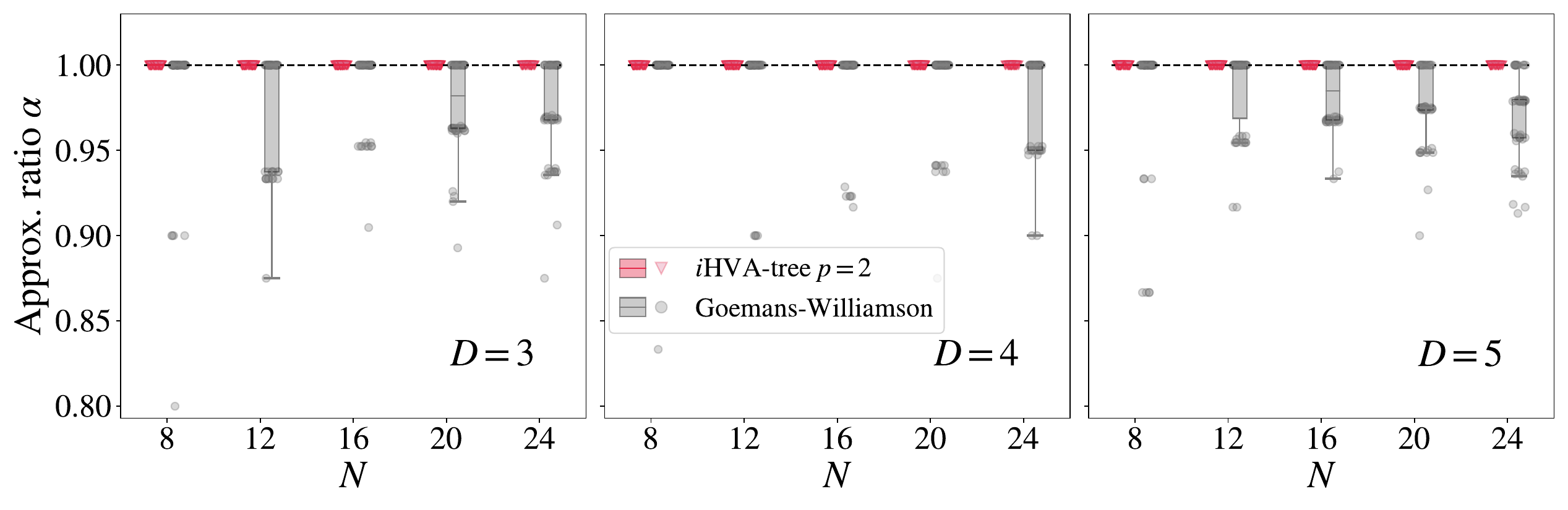}
    \caption{The approximation ratio $\alpha$ of $D$-regular graphs as a function of graph nodes $N$, reached by the two-round $i$HVA-tree and G-W algorithm, where $D=3,4,5$ for each subplot. For each $D$ and $N$, approximation ratios of $50$ random $D$-regular graphs derived by these two algorithms are plotted and marked by red lower-triangle and grey circles, respectively. Compared with the G-W results, $i$HVA-tree can exactly solve MaxCut for all the randomly generated regular graphs with $N$ up to $24$ and $D\leq 5$.}
    \label{fig:qite-tree-3-regular-rounds2}
\end{figure*}

\section{Numerical results}\label{sec:Numerical results}

We have seen that $i$HVA-tree performs better than QAOA ansatz in solving MaxCut of tree graphs. In this section, we numerically demonstrate that the outperformance of $i$HVA-tree can be observed in solving MaxCut of more complicated $D$-regular graphs. We compare the performance of $i$HVA-tree with ma-QAOA ansatz, and further with the classical, polynomial time Goemans-Williamson(G-W) algorithm. The numerical simulations are performed using the noiseless simulator of Qiskit~\cite{Qiskit} and using the \verb|ibm_brisbane| superconducting quantum computer.

\subsection{Simulated results}

We perform numerical simulations using Qiskit noiseless statevector quantum simulator~\cite{Qiskit}. In Fig.~\ref{fig:qite-tree-3-regular}, we plot the approximation ratio as a function of ansatz rounds, with the number of nodes $N \in \{6,8,10,12,14 \}$. For each subplot with a fixed $N$, we generate $50$ random 3-regular graphs~\cite{steger_wormald_1999} with the corresponding number of nodes. The ground state of the Hamiltonian $H_{\mathrm{MC}}$ is prepared using the VQE algorithm, with the classical optimizer \verb|SLSQP|~\cite{2020SciPy-NMeth}. To avoid local minima of the energy landscape as much as possible, for each optimization trajectory , we adopt small constant initialization~\cite{riveradean2021avoiding} for the variational parameters $\bos{\theta}$, where each variational parameter $\theta_{l,ij}$ is chosen independently and uniformly from $[0, 0.001]$. The optimization is performed five times with different initializations for each graph, and the largest approximation ratio $\alpha$ among the five repetitions is selected and plotted. The $i$HVA-tree, $i$HVA-stagger, and ma-QAOA ansatz results are marked by red lower-triangle, green cross, and blue upper-triangle, respectively. The statistical properties of the results are reflected using the box plot~\cite{wiki:Box_plot}, where the middle line of the box denotes the median of the data. For all the number of nodes and the circuit rounds, the performance of the $i$HVA-tree is better than that of the $i$HVA-stagger, and both are better than the ma-QAOA. For $N\leq 14$, approximation ratios achieved by the $i$HVA-tree are all close to one as $p\geq 2$. To achieve the same accuracy, ma-QAOA requires more rounds as $N$ increases, i.e., $p\geq 2,3,4$ for $N=6,8,10$, respectively. This behavior of ma-QAOA is consistent with previous studies ~\cite{mbeng2019quantum}. \looseness-1

For larger graph sizes and regular graphs beyond $D=3$, we perform numerical simulations using $i$HVA-tree with fixed $p=2$. During the optimization, we use Conditional Value at Risk (CVaR) with a confidence level $0.1$ as the objective function, which has been shown to accelerate optimization for combinatorial optimization problems~\cite{Barkoutsos2020improving}. Figure~\ref{fig:qite-tree-3-regular-rounds2} plots the approximation ratios for $50$ graphs and their box plots as a function of graph nodes $N$, where $50$ random $D$-regular graphs are generated for each $N \in \{8,12, \ldots ,24 \}$ and $D \in \{3,4,5\}$. The performance of the classical, polynomial time G-W algorithm on the same test graphs is plotted with grey circles. For each graph, both G-W and VQE with small constant initialization are repeated five times, and the largest approximation ratio is selected and plotted. We see that $i$HVA-tree can exactly solve MaxCut for all the randomly generated regular graphs with $N$ up to $24$ and $D\leq 5$. While the G-W algorithm struggles to achieve the exact solutions for some particular graphs.

\begin{figure}
    \centering
    \includegraphics[width=0.49\textwidth]{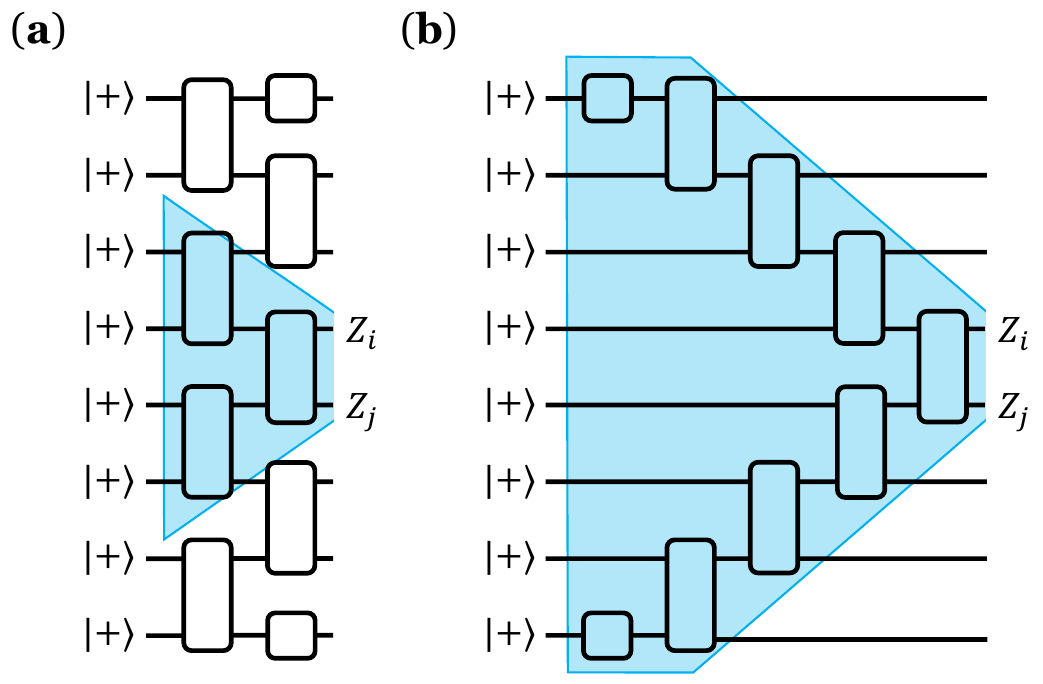}
    \caption{Illustration of the backward light-cone of the observable $Z_i Z_j$ in one round of (\textbf{a}) $i$HVA-stagger or QAOA and (\textbf{b}) $i$HVA-tree on a cycle graph. The two-qubit gate denotes $e^{-i\gamma ZZ/2}$ for QAOA and $e^{-i\theta ZY/2}$ for $i$HVA. Single-qubit gates $e^{-i\beta X/2}$ in QAOA have no impact on the backward light-cone and are not explicitly presented.}
    \label{fig:backward-light-cone}
\end{figure}

The advantage of $i$HVA-tree compared with $i$HVA-stagger and QAOA ansatz can be partially explained by looking at the backward light-cone of an observable $Z_i Z_j$ in these ans\"atze, as shown in Fig.~\ref{fig:backward-light-cone}. The light-cone of the one-round $i$HVA-stagger and QAOA covers only a constant number of qubits, while the light-cone of the $i$HVA-tree covers the whole graph. This means that the one-round $i$HVA-tree is accessible to the global information of the graph, while the $i$HVA-stagger and QAOA are not. The global information is important for the accurate solution of the MaxCut problem~\cite{farhi2020quantum}, and this is also a part of the reason why the MaxCut is hard to solve using classical computers. Additionally, the global backward light-cone indicates that the expectation of $Z_iZ_j$ can not be calculated directly in the Heisenberg picture. On the other hand, for the one-round $i$HVA-stagger and QAOA, the expectation of $Z_iZ_j$ can be calculated efficiently on classical computers by involving a constant number of qubits. From this perspective, $i$HVA-tree could be harder to simulate classically and have more quantum effects involved than $i$HVA-stagger and QAOA ansatz, thus providing higher accuracy than the latter two.

\begin{figure}
    \centering
    \includegraphics[width=0.49\textwidth]{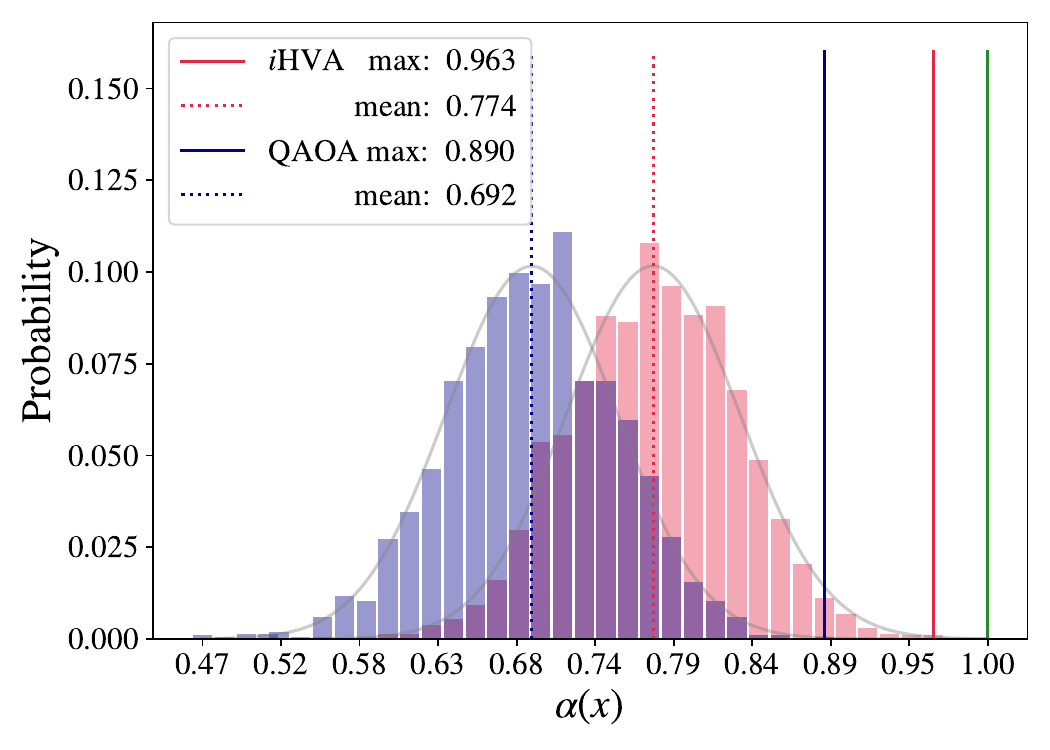}
    \cprotect\caption{Hardware results for the probability distributions of $\alpha(x)$ by 2048 measurements using $i$HVA-tree(red) and QAOA(blue) ansatz. The solid and dashed lines respectively label the position of the max and mean values of the distributions. The green solid line denotes the position of $\alpha(x)=1$. The demonstration is performed using $67$ qubits of \verb|ibm_brisbane|.}
    \label{fig:hardware_results}
\end{figure}

\subsection{Hardware results}\label{sec:Hardware results}
We compare $i$HVA-tree and QAOA ansatz using IBM's quantum hardware to solve the maximum eigenvalue problem. Consider a random weighted Hamiltonian 
\begin{align}
    H_w = \frac{1}{2}\sum_{(i,j)\in \EE}(I-w_{ij}Z_iZ_j),
\end{align}
where $\EE$ is a set of edges of a heavy-hex connectivity graph $G=(\VV,\EE)$ with the number of nodes $N=67$. $w_{ij}$ are randomly chosen to be $\pm 1$. $G$ is tailored to the connectivity of the IBM Eagle-class heavy-hex devices~\cite{IBM_Eagle23,Qiskit}, as shown in Appendix~\ref{app:Setup of the hardware demonstration}. The performance of the variational ansatz is evaluated using an approximation ratio defined by
\begin{align}
    \alpha(x) = \frac{\bra{x}H_w\ket{x} }{\max_{x_0} \bra{x_0}H_w\ket{x_0} },
\end{align}
given a classical bit string $x$ by measuring a quantum state in the Pauli-$Z$ basis. This definition is analogous to the approximation ratio of the MaxCut problem in Eq.~\eqref{eq:approx-ratio-definition}. The exact maximum eigenvalue $\max_{x_0} \bra{x_0}H_w\ket{x_0}$ for the 67-node heavy-hex graph $G$ can be obtained by the greedy algorithm. The greedy algorithm can provide a good approximation to $\max_{x_0} \bra{x_0}H_w\ket{x_0}$ since $G$ has a small number of cycles (See Appendix~\ref{app:Setup of the hardware demonstration}). However, the greedy algorithm does not work for general graphs.

For the hardware demonstrations, we use the equal-angle version of the $i$HVA-tree and the QAOA ansatz with one round, defined by
\begin{equation}
\begin{aligned}
\ket{\phi_I(\theta)} &=
 \prod_{(i,j)\in\EE}e^{-i\theta w_{ij} Z_iY_j/2} \ket{+}^{\otimes N};\\
    \ket{\phi_R(\beta,\gamma)} &=
 \prod_{i\in\VV}e^{-i\beta X_i/2}\prod_{(i,j)\in\EE}e^{-i\gamma w_{ij} Z_iZ_j/2} \ket{+}^{\otimes N},
\end{aligned}
\end{equation}
which have one and two variational parameters to be optimized, respectively. For $i$HVA-tree, the $ZY$ gates are arranged according to the steps provided in Fig.~\ref{fig:qite-tree-ansatz}, and the QAOA gates are arranged to be as shallow as possible~\cite{Bravyi_2020}.

We perform VQE using the above two ans\"atze on the quantum hardware \verb|ibm_brisbane|~\cite{Qiskit}. The $i$HVA-tree circuit performed on the hardware and its layout on the \verb|ibm_brisbane|'s coupling map are shown in Appendix~\ref{app:Setup of the hardware demonstration}. The optimization is performed using the classical optimizer \verb|COBYLA|~\cite{2020SciPy-NMeth} and CVaR as the objective function. Each \verb|COBYLA| optimization trajectory starts using small constant initialization and iterates $20$ steps, and 1024 measurement shots are used to evaluate one CVaR expectation. After the optimization, we take 2048 measurement shots using the optimized variational parameters and calculate their $\alpha(x)$. The probability distributions of $\alpha(x)$ are plotted in Fig.~\ref{fig:hardware_results}(\textbf{b}). We see that both the max and mean $\alpha$ of the $i$HVA-tree are correspondingly larger than the ones of the QAOA ansatz, and the maximum $\alpha$ of $i$HVA-tree reaches 0.963 approximation ratio. These results show the benefit of using $i$HVA-tree over QAOA in solving large-scale combinatorial optimization problems.

When the above two ans\"atze are executed on the hardware, dynamic decoupling with super-Hahn sequence~\cite{Ezzel2023} is used to suppress the decoherence error during the idle periods of the qubits. This technique significantly improves the behavior of the $i$HVA-tree since its qubits have long idle periods due to its tree-like structure.

\section{Absence of Barren Plateau for the constant-round $i$HVA}\label{sec:Trainability of the constant-round $i$HVA}

A variational ansatz with a constant number of rounds is efficient in the number of quantum gates and could be more resilient to decoherence in quantum devices, compared to the ans\"atze with rounds growing with the system size. More importantly, such ans\"atze could be free from barren plateaus (BPs)~\cite{McClean_18}. A variational ansatz with BPs cannot be optimized efficiently due to the exponentially vanishing gradients of its energy landscape. BPs can be diagnosed by calculating the variance of the energy expectation over the variational parameters
\begin{align}
    \mathrm{Var}(\langle \overline{H_{\mathrm{MC}}}\rangle) \equiv \mathrm{Var}_{\bos{\theta}}(\bra{\phi(\bos{\theta})}\frac{H_{\mathrm{MC}}}{E_0} \ket{\phi(\bos{\theta})}),
\end{align}
where $E_0$ is the minimum eigenvalue of the Hamiltonian $ H_{\mathrm{MC}}$, as a normalization factor. If $\mathrm{Var}(\langle \overline{H_{\mathrm{MC}}}\rangle)$ vanishes exponentially with the number of nodes $N$, then the energy landscape of the ansatz is said to exhibit BPs~\cite{Arrasmith_2022}. 

For the constant-round $i$HVA with arbitrary arrangements of the $ZY$ gates, the following theorem holds.
\begin{theorem}\label{theorem2}
For $p$-round $i$HVA in Eq.~\eqref{eq:max-cut-qite-ansatz} solving MaxCut on $D$-regular graph with $N$ nodes, if $p$ is even, the variance of the energy expectation is lower bounded by
\begin{align}
    \mathrm{Var}(\langle \overline{H_{\mathrm{MC}}} \rangle)\geq \frac{DN}{E_0^2 2^{D(p+1)-1}}.
    \label{eq:variance-bound}
\end{align}
\end{theorem}
This theorem is proved by explicitly calculating $\mathrm{Var}(\langle \overline{H_{\mathrm{MC}}} \rangle)$ in the Heisenberg picture, as shown in Appendix~\ref{app:Lack of Barren Plateau}.

According to this theorem, if the degree $D$ is a constant and the ground state energy $E_0$ is of $\OO(\mathrm{poly}(N))$, because the exponent in Eq.~\eqref{eq:variance-bound} does not depend on graph nodes $N$, the variance decays at most polynomially in $N$. Therefore, the constant-round $i$HVA does not exhibit BPs to solve the MaxCut of $D$-regular graphs. This theorem guarantees that the gradient calculated in the previous numerical simulations can be measured efficiently using real quantum devices. 

\begin{figure}[!h]
    \centering
    \includegraphics[width=0.49\textwidth]{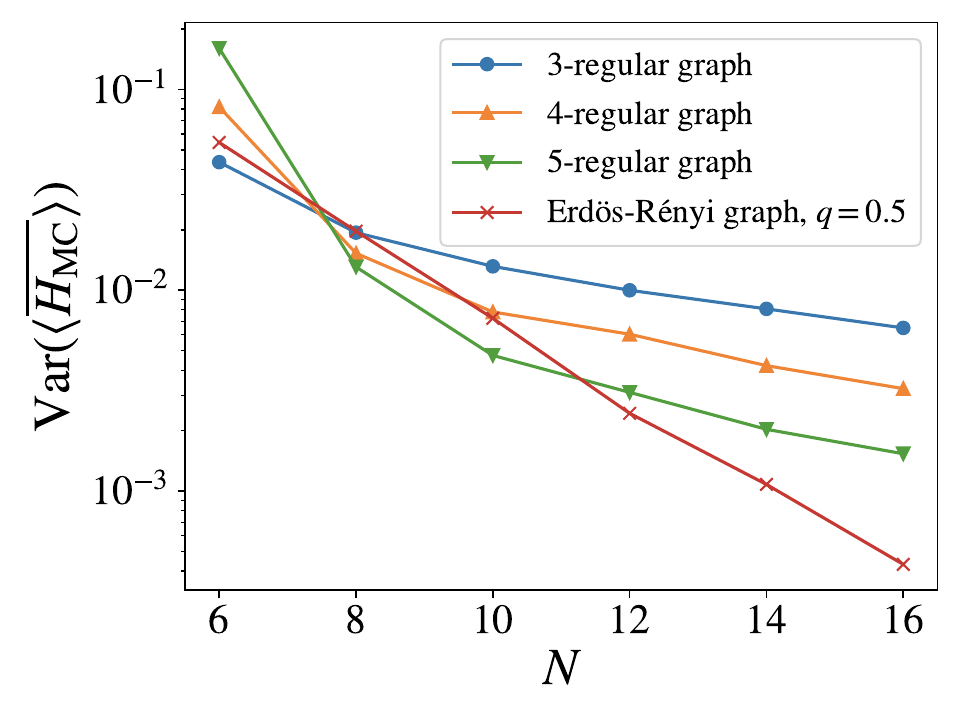}
    \caption{The variance of the normalized energy expectation as a function of the graph nodes $N$. The ansatz used is the two-round $i$HVA-tree. We test $D=3,4,5$-regular graphs and Erd\H{o}s-R\'enyi graphs with $q=0.5$, with results marked by dots, upper-triangle, lower-triangle, and cross, respectively. Within the plot range, the variance decays polynomially with $N$ for regular graphs and exponentially for Erd\H{o}s-R\'enyi graphs. The y-axis is the log scale.}
    \label{fig:qite-tree-ansatz-variance}
\end{figure}

Fig.~\ref{fig:qite-tree-ansatz-variance} depicts the variance of the normalized energy expectation as a function of the graph nodes $N$, where the two-round $i$HVA-tree is used. We randomly generate $50$ graphs, uniformly sample variational parameters $1024$ times from $[0,4\pi]$, and calculate their variance of the energy expectations. Each data point is averaged over the $50$ graphs. It is shown that the variance of $D=3,4,5$-regular graphs decays polynomially with $N$, which is consistent with the theoretical lower bound given in Eq.~\eqref{eq:variance-bound}.

The above theorem can be applied to graphs with more edges than regular graphs with a constant $D$. For example, the all-to-all connected complete graph is an $(N-1)$-regular graph, and Erd\H{o}s-R\'enyi graph connecting each pair of nodes with probability $q$ is effectively a $q(N-1)$-regular graph. For these graphs, the lower bound of the variance decays exponentially with $N$, so the gradient of their constant-round $i$HVA vanishes exponentially with $N$. As numerically demonstrated in Fig.~\ref{fig:qite-tree-ansatz-variance}, for the two-round $i$HVA-tree of Erd\H{o}s-R\'enyi graph ($q=0.5$), the variance decays exponentially with the graph nodes up to $N=16$.

This result shows that the constant round $i$HVA-tree is no panacea. Therefore, special attention has to be paid when using $i$HVA-tree for arbitrary graphs. Although both $i$HVA-tree and QAOA suffer from BPs on certain problems, $i$HVA-tree would still need fewer rounds compared to QAOA. 

\section{Conclusion and Outlook}\label{sec:conclusion}

Common quantum computing approaches to solve the MaxCut problem such as QAOA or quantum annealing suffer from the adiabatic bottleneck that leads to either larger circuit depth or evolution time~\cite{Pelofske_2024, bode_adiabatic_2024, Pelofske_2023}. On the other hand, the evolution time of imaginary time evolution is bounded by the inverse energy gap of the problem Hamiltonian~\cite{Gattringer_2014}. The inverse of the energy gap typically remains constant with system size for many non-critical physical systems, such as the classical Ising chain~\cite{Farhi2000QuantumCB}. Combinatorial optimization problems, including the MaxCut problem, are often modeled using the classical Ising model. This constitutes the motivation to build a variational ansatz based on imaginary time evolution.

In this work, we introduce a variational ansatz that we name the imaginary Hamiltonian variational ansatz ($i$HVA) to solve the combinatorial optimization MaxCut problem. The construction of $i$HVA leverages the bit-flip and time-reversal symmetries of the MaxCut Hamiltonian and the imaginary time evolution. Although $i$HVA is built on the principles of imaginary time evolution, we do not perform imaginary time evolution on quantum circuits, which is distinguished from previous works of QITE algorithm~\cite{McArdle_19,Motta_20,Sun_21,Wang23b}.

The $i$HVA for the MaxCut problem arranges $ZY$ gates utilizing notions of the graph theory. We propose the tree arrangement of $ZY$ gates based on the graph's spanning tree, and show that the MaxCut of arbitrary tree graphs can be achieved exactly by the tree arrangement. Generalizing the tree arrangement of tree graphs onto arbitrary graphs leads to the $i$HVA-tree. We numerically show the advantage of $i$HVA-tree over QAOA in solving the MaxCut problem. The performance of the constant-round $i$HVA-tree is better than the classical Geomans-Williamson algorithm in solving MaxCut on $D$-regular graphs with up to 24 nodes. Additionally, we perform demonstrations using real quantum hardware on a 67-node graph with heavy-hex connectivity, further demonstrating the advantage of $i$HVA-tree over QAOA on the large-scale problem. These results validate our ideas of constructing variational ans\"atze according to imaginary time evolution and oriented spanning tree, or more generally, a directed graph~\cite{AlgorithmDesign2005}. The idea of arranging quantum gates using directed graphs can be adapted to other variational ansatz such as the Hamiltonian variational ansatz (HVA) and the unitary coupled cluster (UCC) ansatz~\cite{Peruzzo:2014}.

Theoretically, the performance guarantees of $i$HVA can be derived similar to the ones given in QAOA~\cite{farhi2014quantum,Wurtz_2021,farhi2020quantum,BassoJoao2022}. The performance guarantees in QAOA are derived based on the locality of the ansatz. However, the $i$HVA-tree proposed in this work is highly non-local, so the method used in QAOA cannot be adapted to the $i$HVA-tree directly. There is no performance guarantee proposed in the literature for non-local variational ans\"atze to our knowledge. While our numerical results demonstrate the benefits of the proposed $i$HVA regarding local minima, saddle points, and the necessary number of rounds, we leave the study of theoretical performance guarantees to future work.

We show that the constant-round $i$HVA-tree of regular graphs does not exhibit BPs. This result has many implications. First, this allows $i$HVA-tree to outperform QAOA on MaxCut problems with regular graphs. Second, this opens up the question of classical simulability of the constant-round $i$HVA-tree based on the recent conjecture by Cerezo et. al~\cite{cerezo_does_2023}. Based on this conjecture, it may be possible that the constant round $i$HVA-tree is classically simulable. For instance, the recently introduced $\mathfrak{g}$-sim method~\cite{goh2023liealgebraic} could be a potential method to simulate constant-round $i$HVA-tree circuits. Such a result would imply the classical easiness of solving MaxCut on $D$-regular graphs. Moreover, this would make the study of $i$HVA-tree on Erd\H{o}s-R\'enyi graphs more valuable, since we demonstrate that constant-round $i$HVA-tree exhibits BPs. In this case, warm start methods such as the one proposed by Chai et al.~\cite{chai2024structureinspired} could be used to support the optimization of $i$HVA.

Finally, the $i$HVA can be constructed for a broader range of quantum systems, such as particle number preserving chemical and condensed matter models, and lattice gauge theories preserving local gauge symmetries. These systems also preserve time-reversal symmetries. Thus, their $i$HVA are distinguished from the commonly employed HVA. One may observe the advantage of the $i$HVA to the HVA in the ground state preparation of these quantum systems. Similarly, Pelofske et al. study the performance of QAOA on higher order Ising models~\cite{Pelofske_2024}. In their results, it can be seen that even short-depth QAOA circuits lead to saddle points and local minima. Employing $i$HVA on higher-order problems has the potential to simplify the optimization processes.

\section*{Acknowledgement}
We thank Maria Demidik, Manuel S. Rudolph and Xiao Yuan for helpful discussions. X.W. and X.F. were supported in part by NSFC of China under Grants No. 12125501, No. 12070131001, and No. 12141501, and National Key Research and Development Program of China under No. 2020YFA0406400. C.T.\ is supported in part by the Helmholtz Association - ``Innopool Project Variational Quantum Computer Simulations (VQCS)''. This work is supported with funds from the Ministry of Science, Research and Culture of the State of Brandenburg within the Centre for Quantum Technologies and Applications (CQTA). This work is funded by the European Union’s Horizon Europe Framework Programme (HORIZON) under the ERA Chair scheme with grant agreement No.\ 101087126. The views expressed here are those of the authors and do not reflect the official policy or position of IBM or the IBM Quantum team.\\

\begin{center}
    \includegraphics[width = 0.10\textwidth]{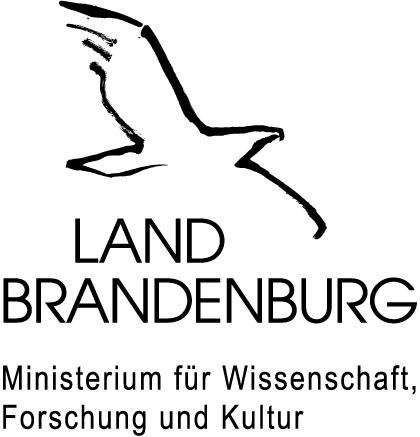}
\end{center}

\onecolumngrid
\appendix

\section{Proof of Theorem \ref{theorem1}}\label{app:exact_cut_of_tree}
The MaxCut of tree graphs can be exactly solved by one round of $i$HVA-tree, as guaranteed by Theorem \ref{theorem1}, which states that
\begin{customthm}{1}
    The MaxCut of arbitrary trees can be achieved by the one-round $i$HVA following the tree arrangement.
\end{customthm}

\begin{figure}
    \centering
    \includegraphics[width=0.48\textwidth]{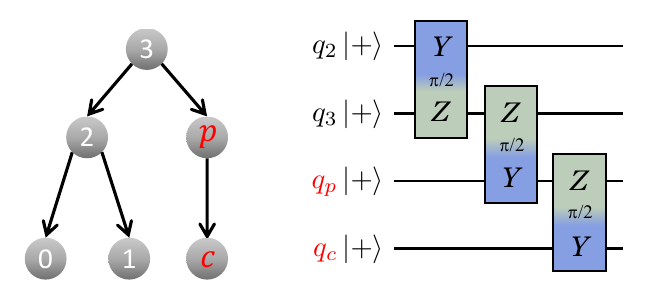}
    \caption{A sketched illustration for the proof of Theorem \ref{theorem1}. The right-most $ZY$ exponential $e^{-i\frac{\pi}{4}Z_p Y_c}$ on the parent node $p$ and the child node $c$ rotates the state $\ket{\phi}=\frac{1}{\sqrt{2}}(\ket{10}_{23}\ket{1}_p+\ket{01}_{23}\ket{0}_p)\ket{+}_c\ket{++}_{01}$ to $\frac{1}{\sqrt{2}}(\ket{10}_{23}\ket{1}_p\ket{0}_c+\ket{01}_{23}\ket{0}_p\ket{1}_c)\ket{++}_{01}$. This circuit is a subcircuit of one shown in Fig.~\ref{fig:tree-arranegment-on-tree}(\textbf{a}).}
    \label{fig:proof-illustration}
\end{figure}

\noindent \textit{proof.} This theorem can be proved by mathematical induction. Our goal is to prepare the ground state of $H_{\mathrm{MC}}$ from the initial state $\ket{+}^{\otimes N}$. In the ground state, for each pair of parent and child nodes of the tree, as defined in Fig.~\ref{fig:graph}(\textbf{a}), their $0/1$ states are opposite. The first step of preparing this ground state is choosing an arbitrary node $p_0 \in\VV$ as the tree's root. Then we choose an arbitrary child node $c_0$ of the root to implement a $ZY$ exponential with parameter $\theta=\pi/2$
\begin{align}
    e^{-i\frac{\pi}{4}Z_{p_0}Y_{c_0}}\ket{++}=\frac{1}{\sqrt{2}}(\ket{01}+\ket{10}),
\end{align}
which leads to opposite $0/1$ states between $p_0$ and $c_0$. For each pair of a parent node $p$ and a child node $c$, the induction hypothesis is that the parent node has been rotated into a component of the GHZ-type state by the upstream $ZY$ gates of the parent node $p$, and the child node remains as the initial state
\begin{align}
    \ket{\phi}\equiv\frac{1}{\sqrt{2}}(\ket{s}\ket{0}_p+\ket{\bar{s}}\ket{1}_p)\ket{+}_c,
\end{align}
where $s$ is a $0/1$ bit string of the upstream qubits of $p$ and $\bar{s}$ is its bit-wise inverse. The downstream qubits of $c$ are in $\ket{+}$ state and are omitted. Implementing a $ZY$ exponential on this state leads to 
\begin{align}
    e^{-i\frac{\pi}{4}Z_p Y_c}\ket{\phi}=\frac{1}{\sqrt{2}}(\ket{s}\ket{0}_p\ket{1}_c+\ket{\bar{s}}\ket{1}_p\ket{0}_c).
    \label{eq:induction}
\end{align}
An example of this implementation is illustrated in Fig.~\ref{fig:proof-illustration}, where the $ZY$ exponential $e^{-i\frac{\pi}{4}Z_p Y_c}$ rotates the state $\ket{\phi}=\frac{1}{\sqrt{2}}(\ket{10}_{23}\ket{1}_p+\ket{01}_{23}\ket{0}_p)\ket{+}_c\ket{++}_{01}$ to $\frac{1}{\sqrt{2}}(\ket{10}_{23}\ket{1}_p\ket{0}_c+\ket{01}_{23}\ket{0}_p\ket{1}_c)\ket{++}_{01}$. Since the resulting state in Eq.~\eqref{eq:induction} is still a GHZ-type state, and the $0/1$ states between $p$ and $c$ are opposite, by induction, $ZY$ gates following the tree arrangement can generate the ground state of $H_{\mathrm{MC}}$ of the tree. Thus, we prove that the targeting ground state can be prepared using the one-round $i$HVA following the tree arrangement. $\qed$

\begin{figure}
    \centering
    \includegraphics[width=0.8\textwidth]{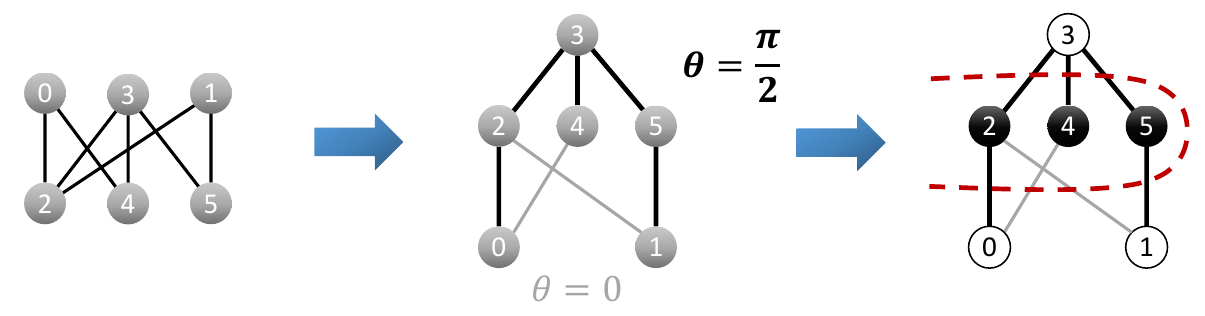}
    \caption{Illustration that the one-round $i$HVA-tree can exactly cut all bipartite graphs. The bipartite graph in the left panel can be exactly cut by cutting its spanning tree (black lines in the middle panel), and the other edges are cut automatically due to the bipartition of the graph (See the right panel). The spanning tree can be cut exactly using the one-round $i$HVA-tree, where the tree edges (black lines in the middle panel) have parameters $\theta=\pi/2$ and the rest edges (grey lines in the middle panel) have parameters $\theta=0$.}
    \label{fig:bipartite-graph}
\end{figure}

A natural corollary of Theorem~\ref{theorem1} is that arbitrary bipartite graphs can be cut exactly using the one-round $i$HVA-tree. The MaxCut of any bipartite graphs can be obtained by cutting its arbitrary spanning trees, as shown in Fig.~\ref{fig:bipartite-graph}. To cut the spanning tree, we set the parameters in the one-round $i$HVA-tree $\theta_{1,ij}=\pi/2$ for edges $(i,j)$ in the spanning tree (black line), and $\theta_{1,i'j'}=0$ for edges $(i',j')$ in the rest of the graph (grey line). In this example, we see that setting all parameters free in $i$HVA-tree helps to improve the solution accuracy of the $i$HVA-tree. On the other hand, if we sets all parameters equal, the bipartite graph cannot be cut exactly using the one-round $i$HVA-tree. This is one of reasons why we choose all free parameters in the construction of $i$HVA.

\section{Circuit depth of $i$HVA-tree}\label{app:circuit-depth-of-iHVA}

\begin{figure}
    \centering
    \includegraphics[width=0.45\textwidth]{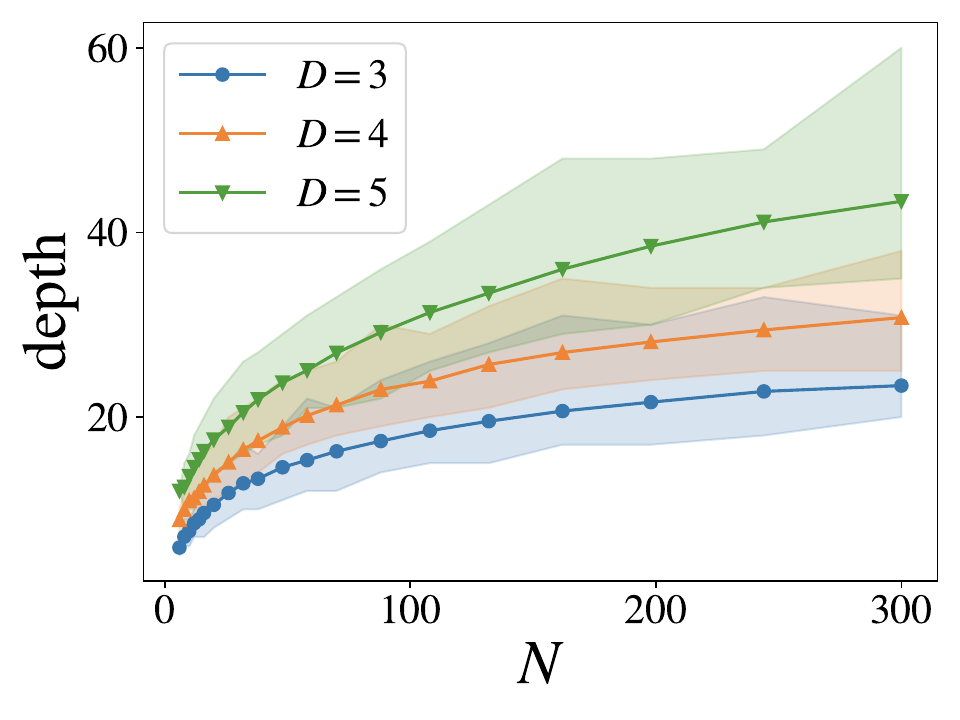}
    \caption{The depth of one round of $i$HVA-tree on $D$-regular graphs as a function of the graph nodes. Results by $D\in\{3,4,5\}$ are marked by dots, upper-triangle and lower-triangle, respectively. Each data point is the average depth, and the colored band denotes the maximum and minimum depth among 200 randomly generated $D$-regular graphs.}
    \label{fig:qite-tree-ansatz-depth}
\end{figure}

The arrangement of $ZY$ gates in $i$HVA-tree leads to the depth of one round $U^{(l)}_{ZY/YZ}$ growing with the number of the graph nodes. In this appendix, we provide  bounds on the depth of $i$HVA-tree of $D$-regular graphs. 

We first consider 2-regular graphs, which are rings, their $i$HVA-tree are ladder-arranged in one round, and the depth grows linearly to the graph nodes $N$. For a $D$-regular graph $G$ with $D>2$, the number of edges is $ND/2$. The depth of one round of $i$HVA-tree can not be larger than the number of edges. Thus we derive an upper bound on the depth of the $p$-round $i$HVA-tree
\begin{align}
    d_p=\OO(pN).
    \label{eq:upper-bound}
\end{align}

On the other hand, the depth of one-round $i$HVA-tree is lower bounded by the height of the graph's spanning tree (See the definition of tree height in Fig.~\ref{fig:graph}(\textbf{b})). Because one descendant edge of a node in the spanning tree increases the circuit depth by one $ZY$ gate depth. The spanning tree of a $D$-regular graph is a $(D-1)$-ary tree, which means that each node has at most $D-1$ child node. Fixing the tree height $h$, a $(D-1)$-ary tree has the maximum number of nodes if every node has $(D-1)$ child nodes. Thus, we have
\begin{align}
    N\leq \sum_{i=0}^{h} (D-1)^i = \frac{(D-1)^{h+1}-1}{D-2}.
\end{align}
As $D>2$, the height of the spanning tree is lower bounded by 
\begin{align}
    h\geq \frac{\log[N(D-2)+1]}{\log(D-1)}-1.
\end{align}
Thus the depth of the $p$-round $i$HVA-tree is lower bounded by
\begin{align}
    d_p=\Omega(p\log N).
    \label{eq:lower-bound}
\end{align}

To verify the derived bounds, we plot the depth of the one-round $i$HVA-tree as a function of $N$, shown in Fig.~\ref{fig:qite-tree-ansatz-depth}. Here, we generate 200 random $D$-regular graphs and count their $i$HVA-tree depth by the depth of the $ZY$ gate. The figure plots the average depth, and the colored band denotes the maximum and minimum depth among the 200 graphs. We see that for $D\in\{3,4,5\}$, the depth grows sublinearly to the number of nodes for the randomly generated graphs. This behavior is consistent with the upper and lower bound given by Eq.~\eqref{eq:upper-bound} and Eq.~\eqref{eq:lower-bound}.

\section{$i$HVA-stagger ansatz}\label{app:$i$HVA-stagger ansatz}
In the main text, we introduce $i$HVA-tree for the MaxCut problem, where the arrangement of $ZY$ gates in $i$HVA is provided explicitly. In this appendix, we provide another arrangement of $i$HVA. This arrangement is inspired by the shallowest arrangement proposed for the QAOA ansatz~\cite{Bravyi_2020}, which makes $i$HVA as shallow as possible. The corresponding $i$HVA with this arrangement is called $i$HVA-stagger.

\begin{figure*}
    \centering
    \includegraphics[width=0.60\textwidth]{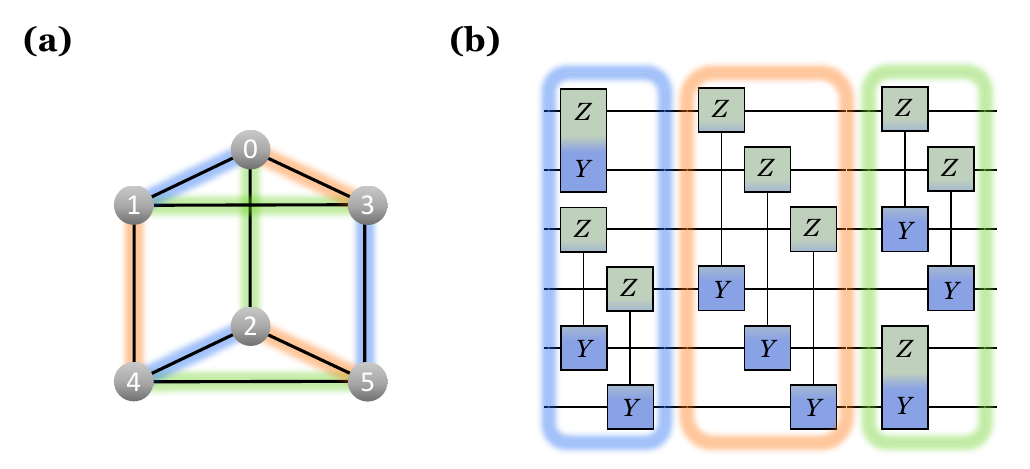}
    \caption{(\textbf{a}) An edge coloring of a 3-regular graph. (\textbf{b}) The subcircuit $U^{(l)}_{ZY}$ following the edge coloring of the 3-regular graph. Two-qubit gates $e^{-i\theta_{l,ij}Z_iY_j/2}$ on edges with the same color can be applied simultaneously.}
    \label{fig:Circuit_MaxCut_3_regular}
\end{figure*}

Consider one round of the $i$HVA 
\begin{align}
    U^{(l)}_{ZY}\equiv \prod_{(i,j)\in\EE}e^{-i\theta_{l,ij}Z_i Y_j/2}
\end{align}
defined on a graph $G=(\VV,\EE)$ with $N$ nodes and a maximum degree $D$. Its $i$HVA-stagger is constructed as follows. First, we make an edge coloring of $G$, which means each edge is assigned a color so that no two incident edges have the same color. For example, Fig.~\ref{fig:Circuit_MaxCut_3_regular}(\textbf{a}) gives an edge coloring for a 3-regular graph with six nodes. According to Vizing's theorem~\cite{Stiebitz2012GraphEC}, there exists an edge coloring utilizing at most $D+1$ colors. Assume $\EE=\EE_1\cup\ldots \cup \EE_{D+1}$ is such an edge coloring. For each color $c\in \{1,\ldots , D+1\}$, we define the following unitary gate
\begin{align}
    U_c^{(l)}=\prod_{(i,j)\in\EE_c} e^{-i\theta_{l,ij}Z_iY_j/2}.
\end{align}
As all the edges in $\EE_c$ are not adjacent, the order of $ZY$ gates in this product is well-defined, and all $ZY$ gates can be realized on quantum devices in parallel. Then, the one-round $i$HVA-stagger is arranged by $ U^{(l)}_{ZY}= U_{D+1}^{(l)}\ldots U_{2}^{(l)} U_{1}^{(l)}$. $U^{(l)}_{ZY}$ of the 3-regular graph in Fig.~\ref{fig:Circuit_MaxCut_3_regular}(\textbf{a}) is shown in Fig.~\ref{fig:Circuit_MaxCut_3_regular}(\textbf{b}). This subcircuit has depth at most $D+1$. Another subcircuit $U^{(l)}_{YZ}$ is constructed by reversing the qubits of $Z$ and $Y$ in $U^{(l)}_{ZY}$, which also has depth at most $D+1$. Therefore, $i$HVA-stagger with $p$ rounds has depth at most $p(D+1)$, which has no dependence on the number of nodes.

For an arbitrary graph, to construct $i$HVA-stagger as shallow as possible, we need to find an edge coloring using the kinds of color as few as possible. The general problem of finding an optimal edge coloring is NP-hard. In practice, we use the greedy coloring algorithm~\cite{Kosowski2008ClassicalCO} to derive an edge coloring with a few kinds of color. 

\begin{figure*}
    \centering
    \includegraphics[width=0.95\textwidth]{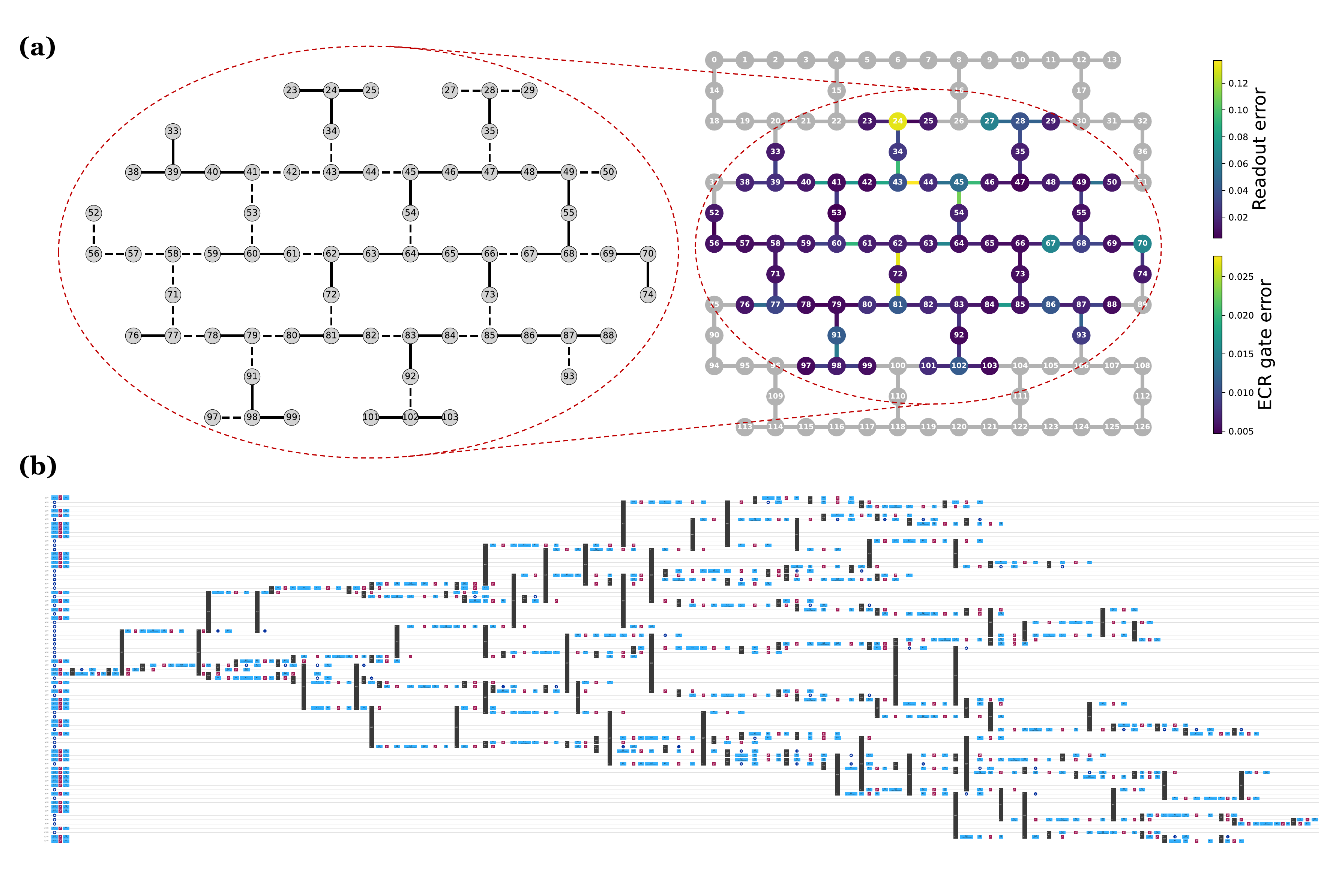}
    \cprotect\caption{(\textbf{a}) The heavy-hex connectivity graph with 67 nodes used in the hardware demonstration (left panel) and its layout on \verb|ibm_brisbane|'s coupling map (right panel). A solid(dashed) line in the graph denotes a weight $w_{ij} = +1(-1)$. The coupling map of \verb|ibm_brisbane| is colored to represent the readout error for each qubit and the two-qubit echoed cross-resonant (ECR) gate error for each qubit connection. (\textbf{b}) The one-round $i$HVA-tree circuit executed on 67 qubits of \verb|ibm_brisbane|. The definition of the quantum gates can be found in the IBM quantum platform~\cite{Qiskit}.}
    \label{fig:graph-circuit}
\end{figure*}

\section{Setup of the hardware demonstration}\label{app:Setup of the hardware demonstration}
In the hardware demonstration shown in Sec.~\ref{sec:Hardware results}, we find the maximum eigenvalue of the random weighted Hamiltonian 
\begin{align}
    H_w = \frac{1}{2}\sum_{(i,j)\in \EE}(I-w_{ij}Z_iZ_j),
\end{align}
where $\EE$ is a set of edges of a heavy-hex connectivity graph $G=(\VV,\EE)$ with $67$ nodes. $w_{ij}$ are edge weights randomly chosen as $\pm 1$, as shown by the solid and dashed lines in Fig.~\ref{fig:graph-circuit}(\textbf{a}). The graph is tailored to the coupling map of the IBM Eagle-class heavy-hex devices \verb|ibm_brisbane|~\cite{IBM_Eagle23} shown in the right panel of Fig.~\ref{fig:graph-circuit}(\textbf{a}). The coupling map is colored to represent the readout error for each qubit and the two-qubit echoed cross-resonant (ECR) gate error for each qubit connection. Other single-qubit properties of \verb|ibm_brisbane| are summarized in Table~\ref{table:qubit-properties}. All hardware data are obtained from the IBM cloud quantum platform~\cite{Qiskit} and more details are available in~\cite{wang_2024}.

The one-round $i$HVA-tree circuit of the heavy-hex connectivity graph executed on \verb|ibm_brisbane| is illustrated in Fig.~\ref{fig:graph-circuit}(\textbf{b}). This figure is generated by the IBM cloud quantum platform~\cite{Qiskit}.
\begin{table}[h]
    \centering
\begin{tabular}{c|cccc}
\hline
 & median & mean & min & max \\
\hline\hline
Frequency (GHz) & 4.91 & $4.90\pm 0.11$ & 4.61 & 5.12 \\
Anharmonicity (MHz) & 308.42 & $308.66\pm 5.38$ & 289.81 & 359.05 \\
$T_1 (\mu s)$ & 222.18 & $218.62\pm 71.64$ & 43.43 & 380.24 \\
$T_2 (\mu s)$ & 142.66 & $151.25\pm 87.70$ & 13.17 & 459.64 \\
\hline
\end{tabular}
    \cprotect\caption{Summary of single-qubit properties of \verb|ibm_brisbane| in the same day of performing the hardware demonstrations.}
    \label{table:qubit-properties}
\end{table}

\section{Proof of Theorem \ref{theorem2}}\label{app:Lack of Barren Plateau}

In this appendix, we prove that the constant-round $i$HVA for the MaxCut problem is free from the Barren Plateaus phenomenon. As explained in the main text, we aim to provide a lower bound on the variance of the Hamiltonian expectation $ \mathrm{Var}(\langle \overline{H_{\mathrm{MC}}}\rangle)$. For convenience, we first consider the $2$-round $i$HVA. Its variance is given by
\begin{align}
    \mathrm{Var}(\langle \overline{H_{\mathrm{MC}}}\rangle) = \mathrm{Var}_{\bos{\theta}}(\bra{\phi_I^{(2)}(\bos{\theta})}\frac{H_{\mathrm{MC}}}{E_0} \ket{\phi_I^{(2)}(\bos{\theta})})=\frac{1}{E_0^2}\Var_{\bos{\theta}}(\sum_{(i,j)\in\EE}\bra{\phi_I^{(2)}(\bos{\theta})}Z_iZ_j \ket{\phi_I^{(2)}(\bos{\theta})}),
    \label{eq:variance of the Hamiltonian expectation}
\end{align}
where 
\begin{align}
    \ket{\phi_I^{(2)}(\bos{\theta})} = U_{YZ}^{(2)}U_{ZY}^{(1)}\ket{+}^{\otimes N},\quad U_{YZ}^{(2)}=\prod_{(i,j)\in\EE}e^{-i\theta_{2,ij}Y_i Z_j/2},\quad U^{(1)}_{ZY}=\prod_{(i,j)\in\EE}e^{-i\theta_{1,ij}Z_i Y_j/2}.
    \label{eq:two-round-QITE}
\end{align}
Then, we have the following lemma.

\begin{lemma}\label{lemma1}
For the $2$-round $i$HVA in Eq.~\eqref{eq:two-round-QITE} solving MaxCut on $D$-regular graph with $N$ nodes, the variance of the energy expectation is lower bounded by
\begin{align}
    \mathrm{Var}(\langle \overline{H_{\mathrm{MC}}} \rangle_2)\geq \frac{DN}{E_0^2 2^{3D-1}}.
\end{align}
\end{lemma}
\noindent \textit{proof.} First, we show that the mean of the Hamiltonian expectation is zero. The mean of the Hamiltonian expectation reads
\begin{align}
    \mathrm{E}_{\bos{\theta}}(\sum_{(i,j)\in\EE}\bra{\phi_I^{(2)}(\bos{\theta})}Z_iZ_j \ket{\phi_I^{(2)}(\bos{\theta})})=\sum_{(i,j)\in\EE} \int \DD\bos{\theta} ~\bra{\phi_I^{(2)}(\bos{\theta})}Z_iZ_j \ket{\phi_I^{(2)}(\bos{\theta})},
    \label{eq:Hamiltonian expectation}
\end{align}
where $\DD\bos{\theta}\equiv \prod_{(i,j)\in\EE}(\frac{\ud\theta_{1,ij}}{2\pi})(\frac{\ud\theta_{2,ij}}{2\pi} )$ is the measure over the circuit parameters. Viewing from the Heisenberg picture, the expectation $\bra{\phi_I^{(2)}(\bos{\theta})}Z_iZ_j \ket{\phi_I^{(2)}(\bos{\theta})}$ can be derived by sequentially conjugating $ZY$ gates on $Z_iZ_j$, for example
\begin{align}
    e^{i\theta Y_iZ_k/2}Z_iZ_j e^{-i\theta Y_iZ_k/2}= \cos\theta Z_iZ_j-\sin\theta X_iZ_jZ_k.
\end{align}
Thus, new Pauli strings such as $X_iZ_jZ_k$ appear in the expression. After conjugating all the $ZY$ gates, it remains a linear combination of Pauli strings. The only Pauli strings contributing to the expectation are those consisting only of Pauli-$X$ letters because
\begin{align}
    \bra{+}^{\otimes N}\sigma \ket{+}^{\otimes N}=\left\{\begin{array}{ll}
        1, & \sigma\in \{I, X_0, X_1, X_0X_1,\ldots ,X_0X_1\ldots X_{N-1}\} \\
        0, & \mathrm{otherwise.}
    \end{array}\right.
    \label{eq:Pauli-X-strings}
\end{align}
Meanwhile, as each of the $ZY$ gates in $ \ket{\phi_I^{(2)}(\bos{\theta})}$ have a free angle, it can be seen that the coefficient of each Pauli string in the linear combination must be a product of $\cos/ \sin$ functions, and each $\cos\theta_{1(2),ij}$ or $\sin\theta_{1(2),ij}$ appears at most once. After the integration $\int \DD\bos{\theta}$, since
\begin{align}
    \int_{0}^{2\pi} \frac{\ud\theta}{2\pi} \cos\theta= \int_{0}^{2\pi} \frac{\ud\theta}{2\pi} \sin\theta=0,
\end{align}
each coefficient in the linear combination vanishes. Thus, the integration $\int \DD\bos{\theta}\bra{\phi_I^{(2)}(\bos{\theta})}Z_iZ_j \ket{\phi_I^{(2)}(\bos{\theta})}$ and the mean of the Hamiltonian expectation in Eq.~\eqref{eq:Hamiltonian expectation} are vanished consequently.

To simplify the calculation of the variance, we assume that all the expectations of $Z_iZ_j$ are mutually independent, i.e., 
\begin{figure*}
    \centering
    \includegraphics[width=0.45\textwidth]{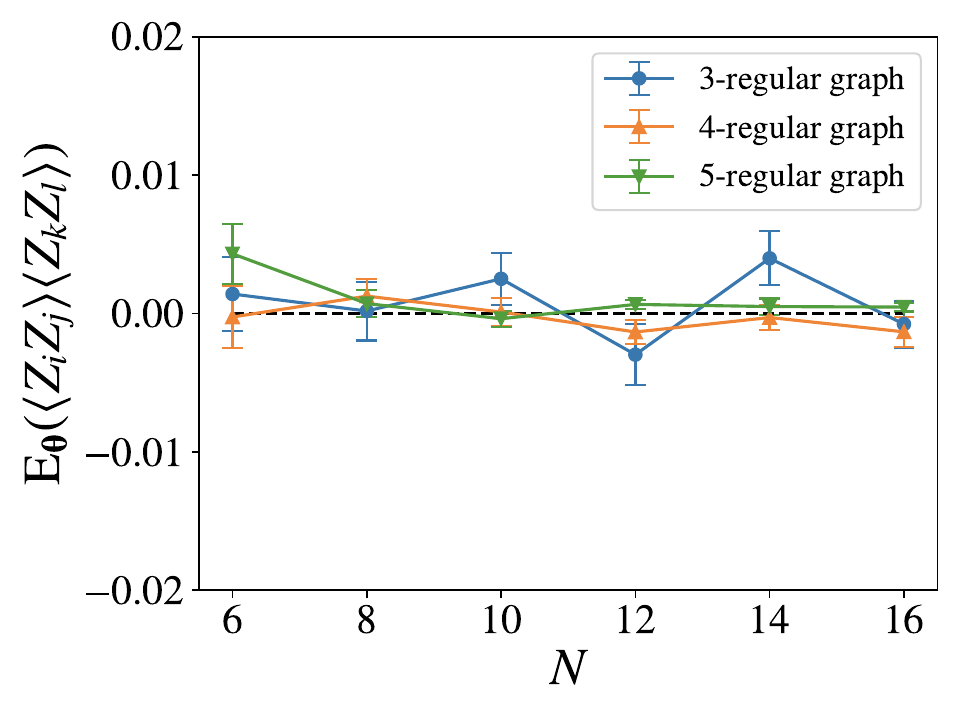}
    \caption{Numerically verification to the assumption in Eq.~\eqref{eq:independence-assumption} that the expectations of observable $Z_iZ_j$ and $Z_kZ_l$ with $i$HVA are mutually independent. Here, a $D$-regular graph with $N$ nodes and the edges $(i,j),(k,l)\in \EE$ are randomly chosen, and the variational parameters $\bos{\theta}$ are uniformly samples 2048 times for each data point. The covariance $\mathrm{E}_{\bos{\theta}}\left(\bra{\phi_I^{(2)}(\bos{\theta})}Z_iZ_j \ket{\phi_I^{(2)}(\bos{\theta})}\bra{\phi_I^{(2)}(\bos{\theta})}Z_k Z_l \ket{\phi_I^{(2)}(\bos{\theta})}\right)$ vanishes for arbitrary $D$-regular graphs with $N$ nodes within the error of statistics. So the left-hand-side of Eq.~\eqref{eq:independence-assumption} is zero with high probability, equal to the theoretical value of the right-hand-side of Eq.~\eqref{eq:independence-assumption}.}
    \label{fig:independence}
\end{figure*}
\begin{equation}
\begin{aligned}
    \mathrm{E}_{\bos{\theta}}\left(\bra{\phi_I^{(2)}(\bos{\theta})}Z_iZ_j \ket{\phi_I^{(2)}(\bos{\theta})}\bra{\phi_I^{(2)}(\bos{\theta})}Z_k Z_l \ket{\phi_I^{(2)}(\bos{\theta})}\right)=&\mathrm{E}_{\bos{\theta}}\left(\bra{\phi_I^{(2)}(\bos{\theta})}Z_iZ_j \ket{\phi_I^{(2)}(\bos{\theta})}\right)\mathrm{E}_{\bos{\theta}}\left(\bra{\phi_I^{(2)}(\bos{\theta})}Z_k Z_l \ket{\phi_I^{(2)}(\bos{\theta})}\right).
    \label{eq:independence-assumption}
\end{aligned}
\end{equation}
in case $(i,j)\neq (k,l)$. This assumption is examined numerically as shown in Fig.~\ref{fig:independence}.
With this assumption, the variance and summation in Eq.~\eqref{eq:variance of the Hamiltonian expectation} can be exchanged, i.e., 
\begin{align}
    \mathrm{Var}(\langle \overline{H_{\mathrm{MC}}}\rangle)=\frac{1}{E_0^2}\sum_{(i,j)\in\EE} \Var_{\bos{\theta}}(\bra{\phi_I^{(2)}(\bos{\theta})}Z_iZ_j \ket{\phi_I^{(2)}(\bos{\theta})}).
    \label{eq:variance of the independent Hamiltonian expectation}
\end{align}


\begin{figure}
    \centering
    \includegraphics[width=0.80\textwidth]{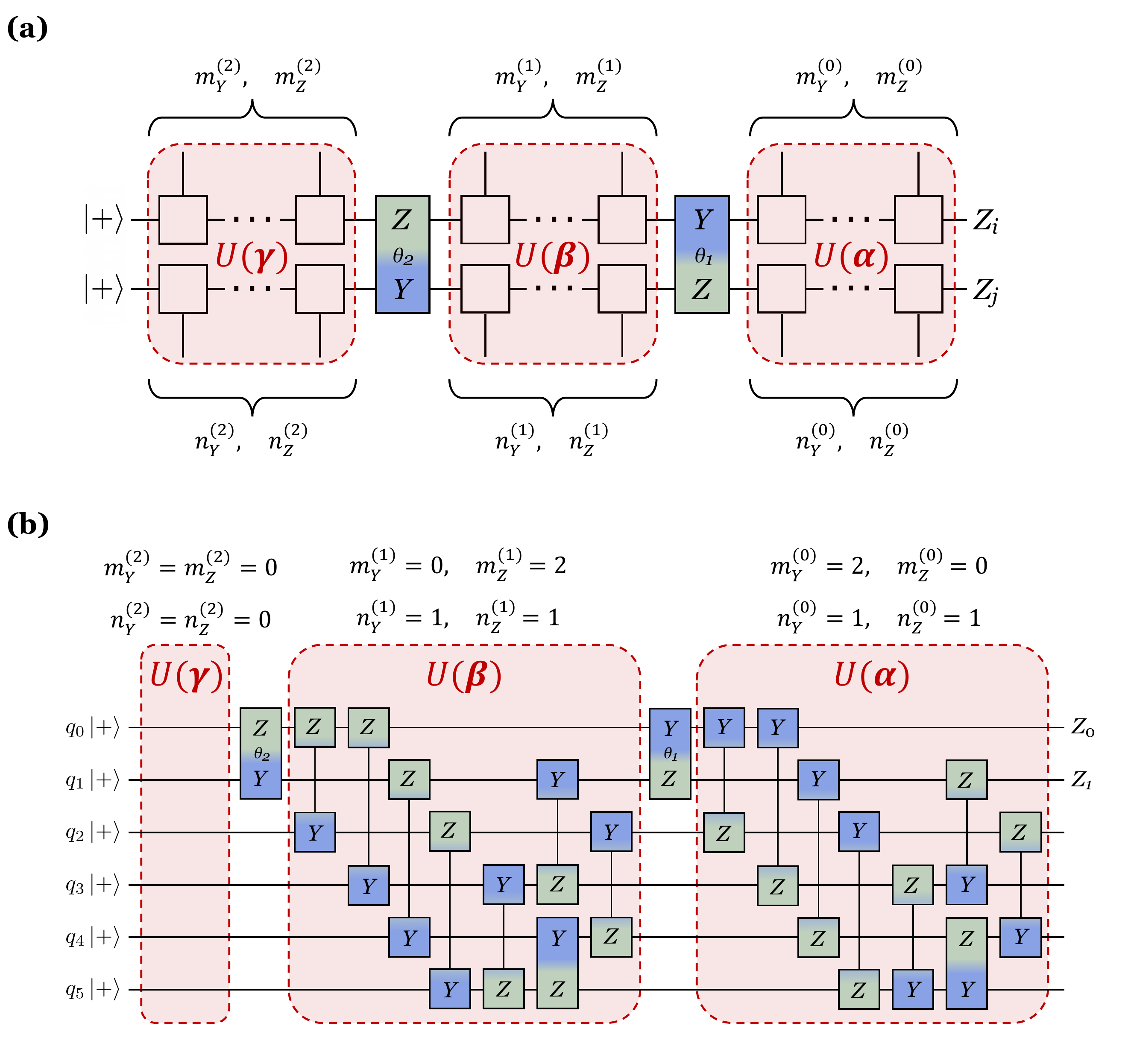}
    \caption{(\textbf{a}) A general structure of $ZY$ gates in the two-round $i$HVA. $U(\bos{\alpha}),U(\bos{\beta}),U(\bos{\gamma})$ are gate blocks of $ZY$ gates. $m_Y^{(i)},m_Z^{(i)},n_Y^{(i)},n_Z^{(i)},i\in\{0,1,2\}$ are the numbers of gates. (\textbf{b}) A concrete example of the gate blocks $U(\bos{\alpha}),U(\bos{\beta}),U(\bos{\gamma})$ and the numbers of gates.}
    \label{fig:QITE_2_structure}
\end{figure}

Next, we provide a rigorous lower bound on the variance $\Var_{\bos{\theta}}(\bra{\phi_I^{(2)}(\bos{\theta})}Z_iZ_j \ket{\phi_I^{(2)}(\bos{\theta})})$. When we calculate the expectation $\bra{\phi_I^{(2)}(\bos{\theta})}Z_iZ_j \ket{\phi_I^{(2)}(\bos{\theta})}$, a general structure of the two-round $i$HVA is shown in Fig.~\ref{fig:QITE_2_structure}(\textbf{a}). In the middle of the circuit, there exists an $e^{-i\theta_{1}Y_i Z_j/2}$ and an $e^{-i\theta_{2}Z_i Y_j/2}$, shown as the colored gates in the figure. As our conclusion has no concern about which qubit is $i$ or $j$, the order of these two gates is irrelevant. Other $ZY$ gates do not connect qubit $i$ and $j$, thus oriented outside the figure. The expectation of $Z_iZ_j$ reads
\begin{align}
    \bra{\phi_I^{(2)}(\bos{\theta})}Z_iZ_j \ket{\phi_I^{(2)}(\bos{\theta})}=\bra{+}^{\otimes N} U^\dagger(\bos{\gamma}) e^{i \theta_2 Z_iY_j/2} U^\dagger(\bos{\beta})e^{i\theta_1 Y_iZ_j/2} U^\dagger(\bos{\alpha})Z_iZ_j U(\bos{\alpha}) e^{-i\theta_1 Y_iZ_j/2} U(\bos{\beta})e^{-i\theta_2 Z_iY_j/2} U(\bos{\gamma})\ket{+}^{\otimes N},
    \label{eq:expectation-first}
\end{align}
where the $ZY$ gate blocks $U(\bos{\alpha}),U(\bos{\beta}),U(\bos{\gamma})$ indicating $ZY$ gates with parameters $\{\bos{\alpha}, \bos{\beta},\bos{\gamma}\}=\{\bos{\theta}\}$, as shown in Fig.~\ref{fig:QITE_2_structure}(\textbf{a}). $m_Y^{(i)},m_Z^{(i)},n_Y^{(i)},n_Z^{(i)},i\in\{0,1,2\}$ in the figure denote the number of gates. For example, $m_Z^{(0)}$ denotes there are $m_Z^{(0)}$ gates in $U(\bos{\alpha})$ connecting qubit $i$ and other qubits like $e^{-i\theta Z_i Y_k/2}$ with $k\in \VV/\{i,j\}$. Fig.~\ref{fig:QITE_2_structure}(\textbf{b}) provides an explicit example of the concrete components of $U(\bos{\alpha}),U(\bos{\beta}),U(\bos{\gamma})$. The corresponding values of $m_Y^{(i)},m_Z^{(i)},n_Y^{(i)},n_Z^{(i)}$ are presented explicitly. These $12$ integers are not independent. Since $i$HVA is constructed for $D$-regular graph, the total number of gates connecting qubit $i$ and other qubits is $2D$ (for two rounds), where $D$ gates are like $e^{-i\theta Z_i Y_k/2}$, and the other $D$ gates are like $e^{-i\theta Y_i Z_k/2}$. Thus, the $12$ gate numbers satisfy the following relations
\begin{align}
    \sum_{i=0}^2 m^{(i)}_Y=\sum_{i=0}^2 m^{(i)}_Z=\sum_{i=0}^2 n^{(i)}_Y=\sum_{i=0}^2 n^{(i)}_Z=D-1.
    \label{eq:mn-constraints}
\end{align}
\begin{figure}
    \centering
    \includegraphics[width=0.88\textwidth]{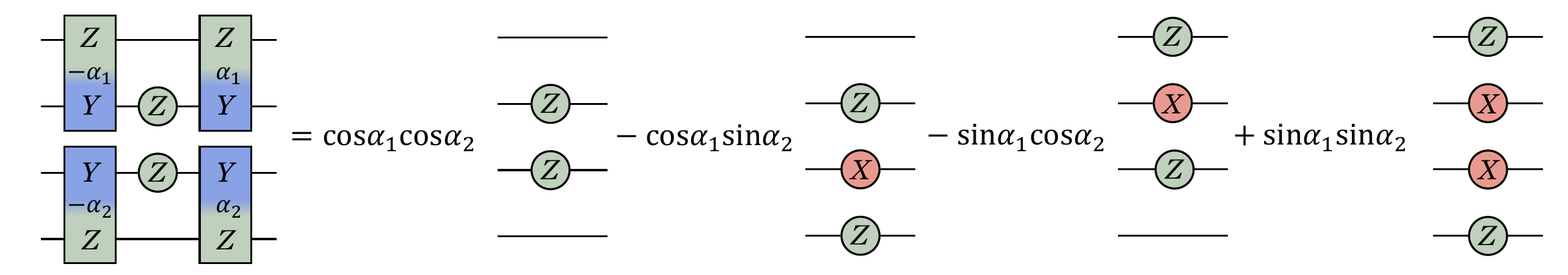}
    \caption{A schematic calculation of the observable $Z_iZ_j$ conjugated by two $ZY$ gates. Each circle denotes a single-qubit Pauli operator and each rectangle denotes a matrix $e^{-i\alpha ZY/2}$ with the value of $\alpha$ shown at the center of the rectangle.}
    \label{fig:QITE_example_circuit}
\end{figure}

With the above notation, we firstly calculate how the rightmost gate block $U(\bos{\alpha})$ in Fig.~\ref{fig:QITE_2_structure}(\textbf{a}) conjugates on the observable $Z_iZ_j$ under Heisenberg picture. An illustration of two gates conjugating on $Z_iZ_j$ is shown in Fig.~\ref{fig:QITE_example_circuit}, where we utilize the relation 
\begin{align}
    e^{i\alpha Y_iZ_k/2}Z_i e^{-i \alpha Y_iZ_k/2}= \cos\alpha Z_i-\sin\alpha X_iZ_k.
\end{align}
It results in a series of Pauli strings with coefficients $\cos\alpha_1\cos\alpha_2,-\cos\alpha_1\sin\alpha_2,-\sin\alpha_1\cos\alpha_2,\sin\alpha_1\sin\alpha_2$. This result can be generalized for $U^{\dagger}(\bos{\alpha})Z_iZ_j U(\bos{\alpha})$. Denote $l_Y=m_Y^{(0)}+n_Y^{(0)}$, which is the number of gates $e^{-i \alpha Y_iZ_k/2},e^{-i \alpha Y_jZ_k/2}$ uncommute with $Z_iZ_j$ in $U(\bos{\alpha})$, the corresponding angles on these $l_Y$ gates are denoted by $\alpha_1\ldots \alpha_{l_Y}$. The observable conjugated by $U(\bos{\alpha})$ is a periodic function of $\bos{\alpha}$, which can be expanded as a high-dimensional Fourier series. The result is a linear combination of $3^{l_Y}$ sub-terms~\cite{you2021exponentially}
\begin{align}
   U^{\dagger}(\bos{\alpha})Z_iZ_j U(\bos{\alpha})=\sum_{\bos{\xi}\in\{0,1,2\}^{l_Y}} \Phi_{\bos{\xi}}(Z_iZ_j) \prod_{l:\xi_l=0}\cos\alpha_l \prod_{l:\xi_l=1}\sin\alpha_l\prod_{l:\xi_l=2} 1,
   \label{eq:UZZU}
\end{align}
where $\xi_l$ is the $l$-th component of the vector $\bos{\xi}\in\{0,1,2\}^{l_Y}$, and
\begin{align}
    \Phi_{\bos{\xi}}(Z_iZ_j)\equiv \sum_{\sigma_{\ii}\in\mathbf{P}^N}d_{\ii}^{(\bos{\xi})} \sigma_{\ii}
\end{align}
is a linear combination of all possible Pauli strings $\sigma_{\ii}\in \mathbf{P}^N=\{I,X,Y,Z\}^{N}$ on $N$ qubits. $d_{\ii}^{(\bos{\xi})}$ are some real coefficients that have no dependence on $\alpha_1,\alpha_2\ldots \alpha_{l_Y}$. The $\bos{\xi}=\bos{0}$ term with coefficients $\prod_{l=1}^{l_Y}\cos\alpha_l$ is explicitly known, i.e., 
\begin{align}
    \Phi_{\bos{0}}(Z_iZ_j)=Z_iZ_j,
\end{align}
This is similar to the results in Fig.~\ref{fig:QITE_example_circuit}. Then, the whole expectation can be derived due to its linearity
\begin{align}
    \bra{\phi_I(\bos{\theta})}Z_iZ_j \ket{\phi_I(\bos{\theta})} = \sum_{\bos{\xi}\in\{0,1,2\}^{l_Y}} a_{\bos{\xi}}\prod_{l:\xi_l=0}\cos\alpha_l \prod_{l:\xi_l=1}\sin\alpha_l\prod_{l:\xi_l=2} 1,
    \label{eq:expectation-2}
\end{align}
where the real coefficients $a_{\bos{\xi}}$ is given by 
\begin{align}
    a_{\bos{\xi}}\equiv &\bra{+}^{\otimes N} U^\dagger(\bos{\gamma}) e^{i \theta_2 Z_iY_j/2} U^\dagger(\bos{\beta})e^{i \theta_1Y_iZ_j/2} \Phi_{\bos{\xi}}(Z_iZ_j) e^{-i \theta_1 Y_iZ_j/2} U(\bos{\beta})e^{-i\theta_2 Z_iY_j/2} U(\bos{\gamma})\ket{+}^{\otimes N}.
\end{align}
As the mean of the expectation vanished, as proved previously, the variance is the integration of the square of the expectation, which reads
\begin{align}
    \Var_{\bos{\theta}}(\bra{\phi_I^{(2)}(\bos{\theta})}Z_iZ_j \ket{\phi_I^{(2)}(\bos{\theta})})=\int\DD\bos{\theta} \bra{\phi_I^{(2)}(\bos{\theta})}Z_iZ_j \ket{\phi_I^{(2)}(\bos{\theta})}^2.
    \label{eq:llllll}
\end{align}
Its integrand given in Eq.~\eqref{eq:expectation-2} is a Fourier series of $\bos{\alpha}$, so the variance can be evaluated using Parseval's equation, which states that if a periodic function can be expanded using Fourier series
\begin{align}
    f(\alpha) = \frac{a_0}{2}+\sum_{n=1}^{\infty} (a_n\cos{n\alpha}+b_n\sin{n\alpha}),
\end{align}
then the integration of the square of $f(\alpha)$ can be derived by 
\begin{align}
    \int_{0}^{2\pi} \frac{\ud \alpha}{2\pi} f(\alpha)^2 = \left(\frac{a_0}{2}\right)^2+\frac{1}{2}\sum_{n=1}^{\infty}(a_n^2+b_n^2),
\end{align}
due to the orthogonality of the Fourier basis. Generalizing this equation to higher dimensions, and integrating out $\bos{\alpha}$ in Eq.~\eqref{eq:llllll}, the variance reads
\begin{equation}
\begin{aligned}
     \Var_{\bos{\theta}}(\bra{\phi_I(\bos{\theta})}Z_iZ_j \ket{\phi_I(\bos{\theta})})=&\int\DD\bos{\theta}/\{\bos{\alpha}\}\sum_{\bos{\xi}\in\{0,1,2\}^{l_Y}}\frac{a_{\bos{\xi}}^2}{2^{\mbox{(the number of 0 and 1 in $\bos{\xi}$) }}}\\
     \geq& \frac{1}{2^{l_Y}}\int\DD\bos{\theta}/\{\bos{\alpha}\} a_{\bos{0}}^2.
     \label{eq:VarZZ}
\end{aligned}
\end{equation}
In the first line, the denominator appears because $\int_{0}^{2\pi} \frac{\ud \theta}{2\pi}\cos^2\theta=\int_{0}^{2\pi} \frac{\ud \theta}{2\pi}\sin^2\theta=\frac{1}{2}$. Each $0$ and $1$ in $\bos{\xi}$ corresponds to a $\cos$ and a $\sin$ function, and leads to one $\frac{1}{2}$. The integration measure $\DD\bos{\theta}/\{\bos{\alpha}\}$ means that the parameters in $U(\bos{\alpha})$ are integrated out. In the second line, we only retain the term $\bos{\xi}=\bos{0}$, whose number of $0$ and $1$ is $l_Y$. So that the factor $1/2^{l_Y}$ appears.

Then, we provide lower bound of $\int\DD\bos{\theta}/\{\bos{\alpha}\} a_{\bos{0}}^2$ using Parseval's equation again. The expression of $a_{\bos{0}}$ can be calculated explicitly
\begin{equation}
\begin{aligned}
    a_{\bos{0}} =& \bra{+}^{\otimes N} U^\dagger(\bos{\gamma}) e^{i \theta_2 Z_iY_j/2} U^\dagger(\bos{\beta})e^{i \theta_1 Y_iZ_j/2} Z_iZ_j e^{-i \theta_1 Y_iZ_j/2} U(\bos{\beta})e^{-i\theta_2 Z_iY_j/2} U(\bos{\gamma})\ket{+}^{\otimes N}\\
    =&\cos\theta_1\bra{+}^{\otimes N} U^\dagger(\bos{\gamma}) e^{i \theta_2 Z_iY_j/2} U^\dagger(\bos{\beta})Z_iZ_jU(\bos{\beta})e^{-i \theta_2 Z_iY_j/2} U(\bos{\gamma})\ket{+}^{\otimes N}\\
    &-\sin\theta_1\bra{+}^{\otimes N} U^\dagger(\bos{\gamma}) e^{i\theta_2 Z_iY_j/2} U^\dagger(\bos{\beta})X_i U(\bos{\beta})e^{-i\theta_2 Z_iY_j/2} U(\bos{\gamma})\ket{+}^{\otimes N}.
\end{aligned}
\end{equation}
These two terms have a similar structure, as we have seen in Eq.~\eqref{eq:expectation-first}, and we can repeat the procedure from Eq.~\eqref{eq:UZZU} to Eq.~\eqref{eq:VarZZ} for both terms. 

Things can be simplified by observing how the factor $1/2^{l_Y}$ appears in Eq.~\eqref{eq:VarZZ}. The exponent $l_Y$ is the number of $ZY$ gates \textit{uncommute} with the observable $Z_iZ_j$. In gate block $U(\bos{\beta})$, the number of gates uncommute with $Z_iZ_j$ and $X_i$ is $m_Y^{(1)}+n_Y^{(1)}$ and $m_Y^{(1)}+m_Z^{(1)}$ respectively. Thus, we come to the lower-bound 

\begin{align}
    \int\DD\bos{\theta}/\{\bos{\alpha}\} a_{\bos{0}}^2\geq   \frac{1}{2^{m_Y^{(1)}+n_Y^{(1)}+1}} \int\DD\bos{\theta}/\{\bos{\alpha},\bos{\beta}\}b_{\bos{0}}^2+\frac{1}{2^{m_Y^{(1)}+m_Z^{(1)}+1}} \int\DD\bos{\theta}/\{\bos{\alpha},\bos{\beta}\}b_{\bos{0}}'^2,
    \label{eq:b1b1}
\end{align}
where 
\begin{equation}
\begin{aligned}
    b_{\bos{0}} &= \bra{+}^{\otimes N} U^\dagger(\bos{\gamma}) e^{i\theta_2 Z_iY_j/2} Z_iZ_je^{-i \theta_2 Z_iY_j/2} U(\bos{\gamma})\ket{+}^{\otimes N};\\
    b_{\bos{0}}' &= \bra{+}^{\otimes N} U^\dagger(\bos{\gamma}) e^{i\theta_2 Z_iY_j/2} X_ie^{-i\theta_2 Z_iY_j/2} U(\bos{\gamma})\ket{+}^{\otimes N}.
\end{aligned}
\end{equation}
Repeating the above procedure and integrating the parameters in $U(\bos{\gamma})$, we derive the lower bound for each term in Eq.~\eqref{eq:b1b1}
\begin{equation}
\begin{aligned}
    \int\DD\bos{\theta}/\{\bos{\alpha},\bos{\beta}\}b_{\bos{0}}^2\geq& \frac{1}{2^{n_Y^{(2)}+n_Z^{(2)}+1}};\\ \int\DD\bos{\theta}/\{\bos{\alpha},\bos{\beta}\}b_{\bos{0}}'^2\geq& \frac{1}{2^{m_Y^{(2)}+m_Z^{(2)}+1}},
    \label{eq:last}
\end{aligned}
\end{equation}
where we have used the fact that only Pauli-$X$ strings contribute to the expectation (See Eq.~\eqref{eq:Pauli-X-strings}).

Collecting Eq.~\eqref{eq:llllll} \eqref{eq:b1b1} and \eqref{eq:last}, the variance of the expectation is lower bounded by
\begin{equation}
\begin{aligned}
    \Var_{\bos{\theta}}(\bra{\phi_I^{(2)}(\bos{\theta})}Z_iZ_j \ket{\phi_I^{(2)}(\bos{\theta})})\geq& \frac{1}{2^{m^{(0)}_Y+n^{(0)}_Y}}(\frac{1}{2^{m_Y^{(1)}+n_Y^{(1)}+1}}\frac{1}{2^{n_Y^{(2)}+n_Z^{(2)}+1}}+\frac{1}{2^{m_Y^{(1)}+m_Z^{(1)}+1}}\frac{1}{2^{m_Y^{(2)}+m_Z^{(2)}+1}} )\\
    =& \frac{1}{2^{D+1+m_Y^{(0)}+m_Y^{(1)}+n_Z^{(2)}}}+\frac{1}{2^{D+1+n_Y^{(0)}+m_Z^{(1)}+m_Z^{(2)}}}\\
    \geq& \frac{1}{2^D} \sqrt{\frac{1}{2^{m_Y^{(0)}+m_Y^{(1)}+n_Z^{(2)}+n_Y^{(0)}+m_Z^{(1)}+m_Z^{(2)}}}} \\
    \geq& \frac{1}{2^{3D-2}}.
    \label{eq:single-variance-lower-bound}
\end{aligned}
\end{equation}

In the second line, we used the constraints in Eq.~\eqref{eq:mn-constraints}. The third line utilizes the basic inequality $a+b\geq 2\sqrt{ab}$. The fourth line is derived by maximizing the exponent $m_Y^{(0)}+m_Y^{(1)}+n_Z^{(2)}+n_Y^{(0)}+m_Z^{(1)}+m_Z^{(2)}$, i.e., choosing $m_Y^{(0)}+m_Y^{(1)}=m_Z^{(1)}+m_Z^{(2)}=n_Z^{(2)}=n_Y^{(0)}=D-1$. Taking the last inequality into Eq.~\eqref{eq:variance of the independent Hamiltonian expectation}, we come to the conclusion 
\begin{align}
     \mathrm{Var}(\langle \overline{H_{\mathrm{MC}}} \rangle) \geq \frac{1}{E_0^2} \frac{DN}{2^{3D-1}},
     \label{eq:app-variance-bound}
\end{align}
where we have used the fact that the number of edges in $D$-regular graph is $DN/2$. $\qed$
  
Now, we prove Theorem~\ref{theorem2} in the main text. Lemma \ref{lemma1} can be generalized to $i$HVA with even $p$ rounds 
\begin{align}
\ket{\phi_I^{(p)}(\bos{\theta}_p)}\equiv U^{(p)}_{YZ}U^{(p-1)}_{ZY} \ldots 
U^{(4)}_{YZ} U^{(3)}_{ZY} U^{(2)}_{YZ}U^{(1)}_{ZY}\ket{+}^{\otimes N}=\mathcal{U}^{(p-2)}\ket{\phi_I^{(2)}(\bos{\theta}_2)},
\label{eq:max-cut-qite-ansatz-app}
\end{align}
where $\{\bos{\theta}_p\}\equiv\{\theta_{1,ij},\ldots, \theta_{p,ij}\}$, $\{\bos{\theta}_2\}\equiv \{\theta_{1,ij}, \theta_{2,ij}\}, (i,j)\in\EE$ and $\mathcal{U}^{(p-2)}\equiv U^{(p)}_{YZ}U^{(p-1)}_{ZY} \ldots 
U^{(4)}_{YZ} U^{(3)}_{ZY}$. The variance of the Hamiltonian expectation reads
\begin{align*}
    \mathrm{Var}(\langle \overline{H_{\mathrm{MC}}} \rangle)  = \frac{1}{E_0^2}\sum_{(i,j)\in\EE} \Var_{\bos{\theta}_p}(\bra{\phi_I^{(p)}(\bos{\theta}_p)}Z_iZ_j \ket{\phi_I^{(p)}(\bos{\theta}_p)})=\frac{1}{E_0^2}\sum_{(i,j)\in\EE} \Var_{\bos{\theta}_p}(\bra{\phi_I^{(2)}(\bos{\theta}_2)}\mathcal{U}^{(p-2)\dagger}Z_iZ_j \mathcal{U}^{(p-2)}\ket{\phi_I^{(2)}(\bos{\theta}_2)}).
\end{align*}
Consider $\mathcal{U}^{(p-2)\dagger}Z_iZ_j \mathcal{U}^{(p-2)}$ with even $p$. For each two-round structure $U^{(k+1)}_{YZ} U^{(k)}_{ZY}$ in $\mathcal{U}^{(p-2)}$, as shown in Fig.~\ref{fig:QITE_2_structure}(\textbf{b}), there are $2D$ $ZY$ gates uncommute with $Z_iZ_j$. These $2D$ $ZY$ gates lead to a production of $2D$ cosine functions as a pre-coefficient of $Z_iZ_j$. Since $\mathcal{U}^{(p-2)}$ has $(p-2)/2$ two-round structures $U^{(k+1)}_{YZ} U^{(k)}_{ZY}$,  $\mathcal{U}^{(p-2)\dagger}Z_iZ_j \mathcal{U}^{(p-2)}$ generates $(p-2)D$ cosine functions as a pre-coefficient of $Z_iZ_j$. Similar to the procedure we used in the proof of Lemma \ref{lemma1}, integrating out the free parameters $\{\bos{\theta}_p\}/\{\bos{\theta}_2\}$ in the variance leads to
\begin{align}
    \Var_{\bos{\theta}_p}(\bra{\phi_I^{(2)}(\bos{\theta}_2)}\mathcal{U}^{(p-2)\dagger}Z_iZ_j \mathcal{U}^{(p-2)}\ket{\phi_I^{(2)}(\bos{\theta}_2)})\geq \frac{1}{2^{(p-2)D}} \Var_{\bos{\theta}_2}(\bra{\phi_I^{(2)}(\bos{\theta}_2)}Z_iZ_j \ket{\phi_I^{(2)}(\bos{\theta}_2)}).
\end{align}

Combining the lower bound of $\Var_{\bos{\theta}_2}(\bra{\phi_I^{(2)}(\bos{\theta}_2)}Z_iZ_j \ket{\phi_I^{(2)}(\bos{\theta}_2)})$ in Lemma \ref{lemma1}, $\mathrm{Var}(\langle \overline{H_{\mathrm{MC}}} \rangle)$ is lower bounded by 
\begin{align}
    \mathrm{Var}(\langle \overline{H_{\mathrm{MC}}} \rangle)\geq \frac{1}{E_0^2} \frac{DN}{2^{3D-1}}\times  \frac{1}{2^{(p-2)D}}=\frac{1}{E_0^2} \frac{DN}{2^{D(p+1)-1}}.
\end{align}
Therefore, we prove Theorem~\ref{theorem2}, and the BP is absent for the constant-round $i$HVA of $D$-regular graphs.

\twocolumngrid

\newpage

\bibliography{main.bib}
\end{document}